\documentclass[useAMS,usenatbib]{mn2e}

\usepackage{longtable}
\usepackage{pdflscape}
\usepackage{graphicx}
\usepackage{pdflscape}
\usepackage{lscape,graphicx,pifont}
\usepackage{color}

\newcommand{\Msun}{\mbox{\,M$_\odot$}}
\newcommand{\Rsun}{\mbox{\,R$_\odot$}}

\newcommand{\vunit}{\mbox{\,km\,s$^{-1}$}}
\newcommand{\mic}{\mbox{$\,\mu$m}}
\newcommand{\pion}[2]{{#1}\,{\sc {#2}}}
\newcommand{\fion}[2]{[{#1}\,{\sc {#2}}]}
\newcommand{\nucl}[2]{\mbox{$^{#1}${#2}}}

\newcommand{\ltsimeq}{\raisebox{-0.6ex}{$\,\stackrel
        {\raisebox{-.2ex}{$\textstyle <$}}{\sim}\,$}}
\newcommand{\gtsimeq}{\raisebox{-0.6ex}{$\,\stackrel
        {\raisebox{-.2ex}{$\textstyle >$}}{\sim}\,$}}

\newcommand{\rnsgr}{\mbox{V3890~Sgr}}

\usepackage{graphicx}	
\usepackage{amsmath}	
\usepackage{amssymb}	

\title[IR spectroscopy of V3890~Sgr]{Infrared spectroscopy of the 
2019 eruption of the recurrent nova V3890~Sgr: separation into
equatorial and polar winds revealed}

\author[]{A. Evans,$^{1}$\thanks{E-mail:a.evans@keele.ac.uk},
T. R. Geballe$^{2}$,
C. E. Woodward$^{3,}$\thanks{Visiting Astronomer at the Infrared Telescope Facility, 
which is operated by the University of Hawaii under contract 80HQTR19D0030 with 
the National Aeronautics and Space Administration.},
D. P. K Banerjee$^{4}$, 
R. D. Gehrz$^{3}$,\newauthor
S. Starrfield$^{5}$,
M. Shahbandeh$^{6,\dag}$\\
\mbox{} \\
$^{1}$Astrophysics Group, Lennard Jones Laboratory, Keele University, Keele, Staffordshire,  ST5 5BG, UK\\ 
$^{2}$Gemini Observatory/NSF's NOIRLab, 670 N. Aohoku Place, Hilo, HI, 96720,
USA\\ 
$^{3}$Minnesota Institute for Astrophysics, School of Physics \& Astronomy,
116 Church Street SE, University of Minnesota, \\
Minneapolis, MN 55455, USA\\ 
$^{4}$Physical Research Laboratory, Navrangpura,  Ahmedabad, Gujarat 
380009, India\\ 
$^{5}$School of Earth and Space Exploration, Arizona State University, 
Box 871404, Tempe, AZ 85287-1404, USA\\ 
$^{6}$Department of Physics, Florida State University, 77 Chieftain Way, Tallahassee, FL 32306-4350, USA
}

\date{Accepted XXX. Received YYY; in original form ZZZ}

\pubyear{2022}

\begin{document}
\label{firstpage}
\pagerange{\pageref{firstpage}--\pageref{lastpage}}
\maketitle

\begin{abstract}
We present infrared spectroscopy of the 2019 eruption of the recurrent 
nova \rnsgr, obtained over the period 5.1--46.3~days after the eruption.
The spectrum of the red giant became more prominent as the flux declined, 
and by day~46.3 dominated the spectrum.
Hydrogen and helium emission lines consisted of a narrow component 
superposed on a broad pedestal. The full width at half maximum of the 
narrow components declined with time $t$ as the eruption progressed, 
as ${t}^{-0.74}$, whereas those of the broad components
remained essentially constant. Conversely, the line fluxes of the narrow
components of Pa\,$\beta$ remained roughly constant, 
while those of the broad components declined by a factor $\sim30$ 
over a period of $\ltsimeq25$~days. The behaviour of the broad components
is consistent with them arising in 
unencumbered fast-flowing ejecta perpendicular to the binary plane,
in material that was ejected in a short $\sim3.3$-day burst.
The narrow components arise in material that encounters 
the accumulated circumstellar material. The outburst spectra were rich 
in coronal lines. There were two coronal line phases, 
one that originated in gas ionised by supersoft X-ray source, the other 
in shocked gas. From the relative fluxes of silicon and sulphur coronal 
lines on day 23.4 -- when the emitting gas was shocked -- we deduce 
that the temperature of the coronal gas was $9.3\times10^5$~K, and
that the abundances are approximately solar.
\end{abstract}




\section{Introduction}
Recurrent novae (RNe) are a subset of cataclysmic
variables. They undergo thermonuclear runaway (TNR) 
eruptions but, unlike classical novae, their 
eruptions recur on timescales of $\sim1-100$~years.
The RNe can be sub-divided into those with short ($\ltsimeq1$~day)
and those with long ($\sim1$~yr) orbital periods
\citep[see][for a review]{evans-aspc}. 
As a consequence of the TNR, up to $10^{-6}$\Msun\ of material,
enhanced in heavy elements (particularly C, N, O, Mg, Al) is
ejected at several 1000s of \vunit\ \citep{anupama08}.

The long-orbital period RNe generally have red giant (RG) 
secondaries which, in common with field RGs, have winds.
When a RN in a system with a RG secondary erupts, 
the ejected material collides with, and shocks, the RG wind, and 
a reverse shock is driven into the ejecta \citep{bode85}. 
This results in strong X-ray and radio emission, and coronal 
line emission in the UV, optical and infrared (IR).
In addition, photoionisation is likely to 
play a part in the production of coronal line emission \citep{munari22}.

We present here a series of near IR (NIR) spectroscopic observations
obtained during the 2019 eruption of the RN \rnsgr. 
Preliminary accounts have been given by \cite{evans19} and
\cite{woodward19a,woodward19b}. A description of various observations 
of \rnsgr\ obtained in quiescence is given by \cite{kaminsky22}.

\begin{figure}
 \includegraphics[width=9cm]{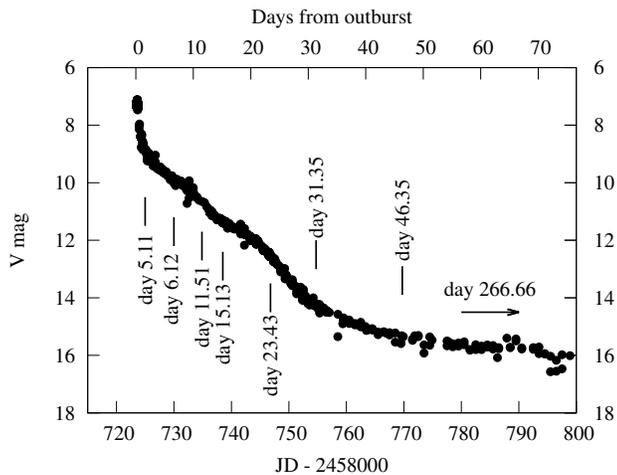}
\caption{$V$-band light curve of \rnsgr; data from the 
AAVSO archive. Times of the observations reported here are 
indicated.
\label{AAVSO_LC}}
\end{figure}

\section{Properties of \rnsgr}

\subsection{The binary}

\rnsgr\ underwent eruptions in 1962 and 1990 \citep[e.g.][]{anupama08}.
Its latest eruption was on 2019 Aug 27.87 
\citep[JD~2458723.37;][]{pereira}. We take this to define our zero of time,
$t=0$. It has been reported to have an orbital period of $519.7\pm0.3$~days 
and semi-major axis 362\Rsun\ \citep{schaefer09}, but a more recent
analysis, using extensive spectroscopic and photometric data 
\citep{mikolajewska21}, gives an orbital period of 747.6~days. 
\citeauthor{mikolajewska21} give the mass of the white dwarf (WD) 
as $M_{\rm WD} \simeq 1.35\pm 0.13$\Msun, consistent with the view that 
the masses of the WD components in RNe are close to the Chandrasekhar 
limit \citep*{starrfield88,anupama08}.

The RG in \rnsgr\ has a mass $M_{\rm RG} \simeq 1.05\pm0.11$\Msun, and 
a M5~III classification \citep{anupama08}. This implies that, as in
other examples of this sub-class of RNe, the material ejected during
the TNR will encounter the RG wind, driving a shock into the wind,
and a reverse shock into the ejecta \citep{bode85}.
The orbital inclination is $\sim67-69^\circ$\citep{mikolajewska21};
we assume $68^\circ$ here.

The properties of \rnsgr\ in quiescence have been described in detail
by \cite{kaminsky22}. For the RG, they found small overabundances
of both carbon and sodium relative to their solar values, while the 
abundances of oxygen and silicon are solar. The \nucl{12}{C}/\nucl{13}{C} 
ratio was found to be $\sim25$, similar to that found in the RG
components of other RNe.

\cite{kaminsky22} also found that a cool ($\sim400$~K) component 
is required to fit the continuum. This is clearly due to dust in the 
\rnsgr\ system. Silicate dust has been found in the environment of the RN RS~Oph
\citep{evans07c,woodward08,rushton22}, which is likely distributed in
a torus around the binary; however there is no evidence for silicates in the 
environment of \rnsgr\ \citep{kaminsky22}.

Based on the pulsation of the RG, and on the assumption that the RG 
fills its Roche lobe, \cite{mikolajewska21} determined a distance of
$D \simeq 9$~kpc.

\subsection{Interstellar reddening}
\cite{munari19a} measured the equivalent widths of interstellar
\pion{Na}{i} and \pion{K}{i} absorption lines seen in high 
resolution optical spectra. They found $E(B-V)=0.56$.
They also noted that, as the intrinsic value of $(B-V)$ for novae 
2~magnitudes from maximum is $(B-V)_0=0.02\pm0.04$ \citep{vdB}, its 
early value of $(B-V)=0.67$ is consistent with $E(B-V)=0.67\pm0.2$.
A thorough review of the reddening to \rnsgr\ was given by 
\cite{page20}, who found values between $E(B-V)=0.48$ and 0.59.
\cite{kaminsky22}, in their anlysis of the IR spectrum, found that a
value $E(B-V)=0.40$ gives the best fit. We take $E(B-V)=0.5$ 
here.

\subsection{The 2019 eruption}
\label{2019}
The most recent RN eruption of \rnsgr\ was reported by \cite{pereira}, and
an optical spectrum was obtained within hours of outburst
by \cite{strader19}. Early NIR spectroscopy was reported by
\cite{rudy19}, who found that the emission lines were broad,
with full widths at zero intensity (FWZI) of $\sim10000$\vunit.
\cite{munari19b} described the early optical emission lines
as roughly triangular in profile, but containing a narrower
($\sim500$\vunit) component as the lines weakened; they also noted 
that the FWZI of H$\alpha$ was $\sim8300$\vunit\ on August 29
(close to visual maximum), and $\sim7900$\vunit\ on 
September 11 ($\sim14$~days after outburst), suggesting that 
little deceleration had taken place during that early interval. 
\cite{munari19b} also noted the presence of very sharp 
emission features, superimposed on the broad lines, 
which had disappeared by day~3.4. They suggested that these 
components arose in the flash-ionised RG wind, their 
disappearance being due to recombination, suggesting an 
electron density of $4\times10^7$~cm$^{-3}$ in the wind. 
\rnsgr\ was detected in the radio (1.28~GHz) very early in the
eruption by \cite{nyamai19}.

Extensive X-ray observations of the 2019 eruption were described by 
\cite{orio20}, \cite{page20} and \cite{ness22}. \citeauthor{page20} reported
that the supersoft source (SSS) phase began around day~8.6,
and persisted until day~26. They found that the temperature of 
the supersoft source rose to $T\sim1.2\times10^6$~K  ($kT\sim100$~eV) 
between days 9.24 and 13.28, and that the spectrum steadily softened
thereafter, up to day $\sim26$, when the effective blackbody temperature
was $T\sim4.6\times10^5$~K ($kT\simeq40$~eV). The SSS phase in 
\rnsgr\ occurred much earlier in the eruption, and ended a much sooner,
than was the case during either the 2006 or 2021 eruptions 
of the RN RS~Oph \citep{orio20,page20}. 

\rnsgr\ is one of the growing number of novae, 
both classical and recurrent, to have been detected in $\gamma$-rays 
during eruption \citep*{buson19}.
In systems having RG secondaries, the observed
high-energy $\gamma$-rays probably arise as a result of collision 
between the RN ejecta and the RG wind, although in classical novae 
the most likely interpretation is shocks internal to the ejecta 
\citep*[see][for a thorough review; also \cite{aydi20}]{chomiuk21}.

The $V$-band light curve from the AASVO 
archive\footnote{https://www.aavso.org/}
is shown in Fig.~\ref{AAVSO_LC}.

\section{Observations}

\subsection{Gemini Observatory}

NIR spectra of \rnsgr\ were obtained at both the Gemini South 
telescope \citep[with the facility spectrometer 
FLAMINGOS-2;][]{eikenberry12} and 
the Frederick C. Gillett Gemini North telescope \citep[with the facility
spectrometer GNIRS;][]{elias06}. Pertinent observational parameters are given in 
Table~\ref{log}. All observations were obtained in the standard 
stare/nod-along-slit mode. 
Unlike most GNIRS cross-dispersed spectra, these GNIRS spectra were 
obtained with the $111\ell\,$mm$^{-1}$ grating in order to obtain 
higher spectral resolution, and required three wavelength settings
of that grating to cover the entire 0.8-2.5\mic\ interval. 

For the purpose of flux calibration and 
removal of telluric absorption lines, a nearby early A-type dwarf
star was observed either immediately before or immediately after \rnsgr. 
The airmass difference between the science target and the telluric standard
was in all cases $<0.13$.
Both \rnsgr\ and the telluric standards were observed with the 
slit oriented at the average parallactic angle. 

Data reduction utilising both {\sc iraf}\footnote{{\sc iraf} is distributed 
by the National Optical Astronomy Observatories, which are operated by
the Association of Universities for Research in Astronomy, Inc., under
cooperative agreement with the National Science Foundation.}
\citep{tody1,tody2}
and {\sc figaro} \citep{figaro}, employed
standard procedures of spectrum extraction, spike removal, wavelength
calibration (using spectra of argon arc lamps), removal of \pion{H}{i} lines from the
spectra of the standard star, cross correlating the spectra of 
\rnsgr\ and the telluric standard star and shifting the spectrum of 
the former to align it with the spectra of the standard star, and 
ratioing. The resolutions of the GNIRS spectra are roughly an order
of magnitude higher than those of the FLAMINGOS-2 spectra. 
In the figures presented here the GNIRS spectra are sampled roughly
an order of magnitude more finely than the FLAMINGOS-2 spectra 
(0.00010--0.00020\mic\ versus 0.0010--0.0015\mic).  

\subsection{IRTF}

\rnsgr\ was observed with the NASA IRTF 3.2-m telescope on multiple epochs with
the facility IR spectrograph SpeX \citep{2003PASP..115..362R} in both 
SXD (short-crossed dispersed) and LXD (long-short crossed dispersed) modes,
using slits matched to the seeing conditions, nodding the source at two positions
(ABBA mode) in the slit set to the parallactic angle on the sky appropriate for
the time of observations. Standard flat-field and calibration lamp (used to 
determine the wavelengths) spectra were obtained at the point position of the
target. All spectra (\rnsgr\ and the associated telluric stars) were reduced
using the IRTF IDL-based Spextool package 
\citep[version v4.1;][]{2004PASP..116..362C} following standard IR techniques.
Optimal point-source spectral extraction of the two-dimensional raw data frames
was used. The aperture profiles were traced, and spectra extracted with 
parameters that typically used a Point Spread Function (PSF) radius and 
Aperture radius $=2\farcs2$ and $1\farcs0$ respectively. Background subtraction
was enabled with the background sampling region starting at 
$2\farcs2$ (just outside the PSF radius) with a sampling width of $2\farcs0$. 
The background was generated by a polynomial fit of degree $= 1$ to data in
the latter sampling region. The individual
extracted spectra were then scaled (using order 3 for SXD and order 6 for LXD
observations) and combined, after correcting the spectral shape of the orders,
using a robust weighted mean technique with threshold value of 0.8 
\citep[as described in][]{2004PASP..116..362C}.  Flux calibration of 
the \rnsgr\ spectra was achieved using spectral type AOV telluric stars with
known $B$ and $V$ magnitudes, following the methods described in
\citet*{2003PASP..115..389V}. 
The deconvolution method was used for the SXD
observations, using the hydrogen Pa\,$\gamma$ (1.09411\mic) absorption feature
in the continuum normalized spectra to construct the kernel relative
to $\alpha$~Lyrae (Vega), with typical residuals having maximum deviations
of $\ltsimeq3$\% and RMS deviation of $\ltsimeq1.6$\%.  
For the LXD observations, the convolution kernel was established from the
arc lamp lines (Instrument Profile approach).  
Residual stellar H lines in the
telluric spectra were removed through inspection and interpolation
to produce telluric spectra containing only atmospheric absorption spectra.  
Lastly, each order was inspected to determine whether residual wavelength 
shifts were necessary to avoid introducing noise or artifacts when the
\rnsgr\ spectra were divided by the telluric spectra.

Table~\ref{log} provides details of the Gemini and IRTF SpeX observations.
The times of the IR spectroscopic observations are shown in Fig.~\ref{AAVSO_LC}.

\section{Overview of the spectra}

\begin{figure*}
\centering
\includegraphics[width=18cm]{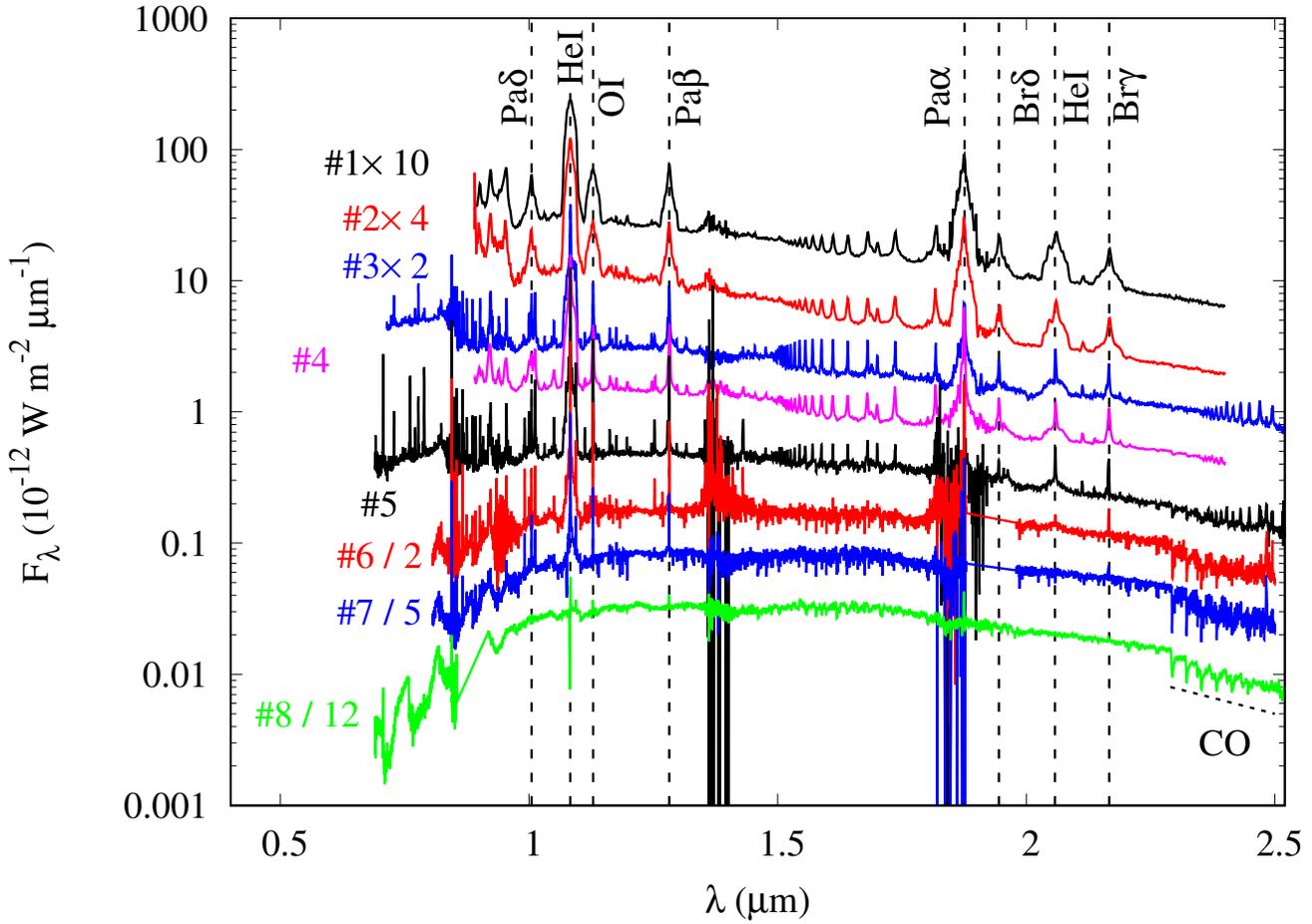}
\caption{Evolution of the IR spectrum.
See Table~\ref{log} for key. Individual spectra have been 
multiplied/divided by the factors indicated for clarity.
Stronger emission features are identified, as are the CO
first overtone bands.
See Table~\ref{coronals} and Fig.~\ref{all_1} for further details.
Note the noisiness of some of the spectra around 1.4\mic\ and 1.8\mic\
due to the poor atmospheric transmission at these wavelengths.
\label{plot_all}}
\end{figure*}

The entire dataset is displayed in Fig.~\ref{plot_all}.
Some key features are summarised here, and are discussed in detail in
the following subsections.
\begin{enumerate}
 \item The hydrogen and helium emission lines had narrow components,
 superposed on broad pedestals.
 \item By day 15.13, the first overtone CO bands, band-head
 at 2.29\mic, were discernible in absorption;
 these must originate in the RG photosphere. They became progressively more
 prominent with time as the contribution from the ejected gas declined.
 \item Coronal lines were present from the earliest spectrum (day~5.11);
 unlike the H and He lines, they were narrow from the outset. 
 \item On days 11.51 and 23.43 (when the spectra included wavelengths
 below 0.8\mic), the Paschen (0.82\mic) and Brackett (1.46\mic) discontinuities were
 present and substantial, indicative of a relatively cool gas.
\end{enumerate}

\subsection{The continuum}

As the continuum flux declined during the outbust, the contribution 
from the RG component became increasingly prominent. 
In order to facilitate the analysis of the nebular spectrum,
the spectrum of the RG, obtained on 2020 May 20 
\citep[day 266.66; described in][]{kaminsky22}, was subtracted 
from the data; there were no coronal lines on day 266.66.
The spectrum from 2019 September 8 (day~11.51), dereddened and 
with the contribution from the RG removed, is shown in the 
left panel of Fig.~\ref{ff-bf};
the continuum is fitted with a pure hydrogen nebular continuum 
at electron temperature $\sim7000$~K and electron density $n_e=10^7$~cm$^{-3}$
(although the shape of the nebular continuum is not sensitive to $n_e$). 

The right panel of this figure gives the same information, but for
2019 September 20 (day~23.43). In this case the fit, for a 6000~K nebular
continuum, is not as good, but the large Paschen jump indicates a
relatively low temperature gas. Such a low 
temperature is required to fit the Paschen and Brackett discontinuities
at $\sim0.82$\mic\ and 1.46\mic\ respectively.

As noted above, coronal lines, indicative of hot ($\sim10^5$~K),
low density gas, are already present in the earliest spectrum. The 
cool (7000~K) gas observed on day~11.51 is unlikely to originate in the 
RG wind, flash-ionised by the ultraviolet pulse from the eruption,
because the flash-ionised component had recombined by day~3.4 and had 
disappeared from the spectrum
\citep[][see also Section~\ref{2019}]{munari19b}.
This flash-ionised component was not present in our first observation 
(day~5.11). However, with the spectral resolutions at our disposal 
at that time (see Table~\ref{log})
our observations would not have detected such narrow features.
The presence of substantial Paschen and Brackett discontinuities
was recorded on day 11.81 of the 2006 eruption of the RN RS~Oph 
\citep{evans07a}, but their explanation, in terms of flash-ionisation,
may need reassessment.

\begin{landscape}
 
\begin{table*}
\centering
\caption{Log of infrared observations. ``Day'' is number of days
since the start of the eruption on 2019 Aug 27.87 (JD 2458723.37).}
\label{log}
\begin{tabular}{ccrccccccccc} \hline
UT Date       & JD --    & Day  &Facility  & Instrument& Airmass & Wavelength  & Int.& Slit& Res'n & Telluric & Key to\\
{\small YYYY-MM-DD.dd} & 2458000       & &          &           &         & range (\mic)&time (s)  & width ($''$) & $\lambda/\Delta\lambda$ &  standard& Fig.~\ref{plot_all}\\ \hline
2019-09-01.98 &728.48& 5.11  &Gem-S&F-2 $J\!H$&  1.03  & 0.89--2.40 &8 &0.36& 200--1200  & HIP97692 &\#1\\
&        &        &          &  ~~~~    $H\!K$    & 1.02  &       &8  &0.36&  300--1200  & HIP97692 & \#1\\
2019-09-02.99 &  729.49  & 6.12    & Gem-S & F-2  $J\!H$ & 1.03     &0.89--2.40  &  8 &0.18&  200-1400 & HD152602 & \#2\\
&        &        & &  ~~~~ $H\!K$ &   1.02     &    & 8 &0.18&  300--1500 & HD152602 & \#2 \\
2019-09-08.38 & 734.88 & 11.51  & IRTF     &  SpeX       &   2.33     & 0.71--2.56 & 717  && 2000 & HIP93901 & \#3\\
2019-09-12.00 &  738.50 & 15.13       & Gem-S & F-2 $J\!H$  &  1.01 & 0.89--2.4   & 24&0.36&  200--1200     & HIP85607 & \#4\\
             &         &             &       & F-2 $H\!K$  &  1.01 &              & 16&0.36&  200--1200     & HIP85607 & \#4\\
2019-09-20.30 & 746.80 &  23.43 & IRTF     &  SpeX   &    1.74    & 0.69--2.56  & 20 &  &2000 & HIP93691  & \#5\\
2019-09-28.22 & 754.72 &  31.35  & Gem-N &  GNIRS  &  1.44      & 0.8--2.5 & 600 & 0.3& 6000  &HIP113673 & \#6\\
2019-10-13.22 & 769.72 &  46.35 & Gem-N &     GNIRS     &  1.57      & 0.8--2.5  & 600 & 0.3&6000 & HIP94510 & \#7\\ 
2020-05-20.53 & 990.03 & 266.66 & IRTF     & SpeX    & 1.39   & 0.69--2.56 & 1618 &0.8& 750  & HD168707 & \#8\\\hline
\end{tabular}
\end{table*}

\end{landscape}

\subsection{Emission lines}

\subsubsection{H and He lines}
\label{hwhm}

Selected segments of the spectra, in the regions of particular
emission lines, are shown in Fig.~\ref{2mic}.
It is evident from Figs~\ref{plot_all} and \ref{2mic} that some emission 
lines are broad on day~5.11, with narrow cores; the narrow components 
become progressively narrower with time. This is probably because 
ejected material is colliding with, and is decelerated by, the RG wind 
and any circumbinary material in the orbital plane that remains
from previous eruptions and the common envelope phase. However, as 
discussed in Section~\ref{disc}, the picture is more complex than this.

\begin{figure*}[h]
\includegraphics[width=7.cm]{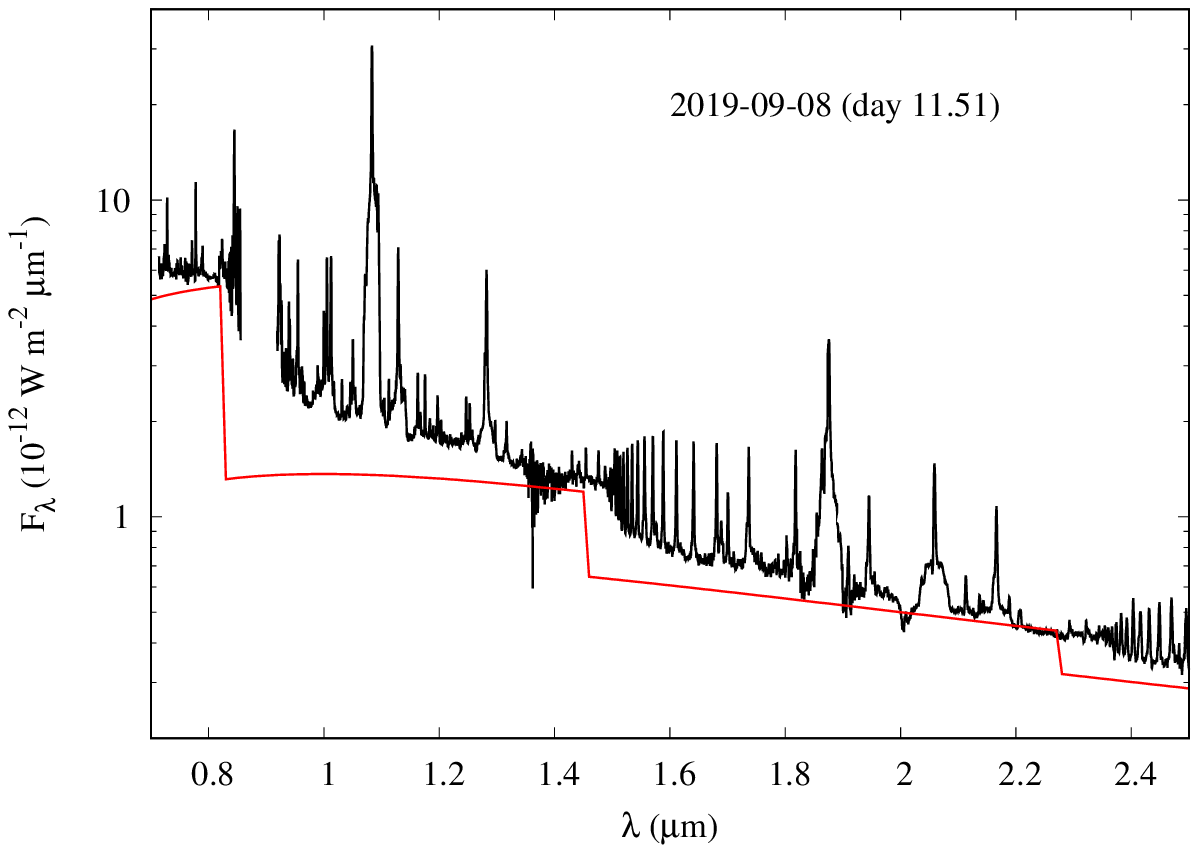}
\includegraphics[width=7.cm]{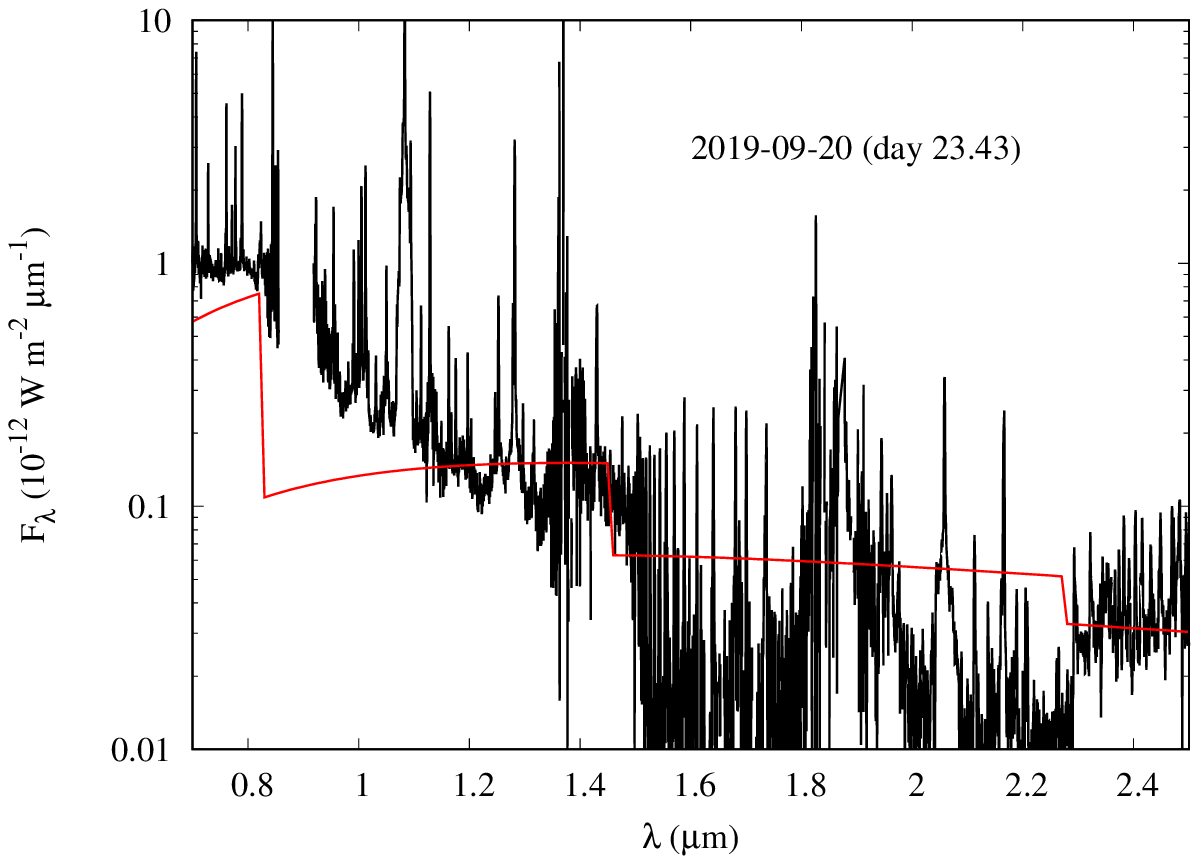}
\caption{Top: Spectrum on 2019 September 8, dereddened and with RG 
contribution subtracted; the red curve is a 7000~K nebular 
continuum, normalised at 2.24\mic.
Bottom: As top but on 2019 September 20 and with 6000~K nebular 
continuum, normalised at 2.4\mic.\label{ff-bf}}
\end{figure*}

 \begin{figure*}
 \includegraphics[width=7.cm]{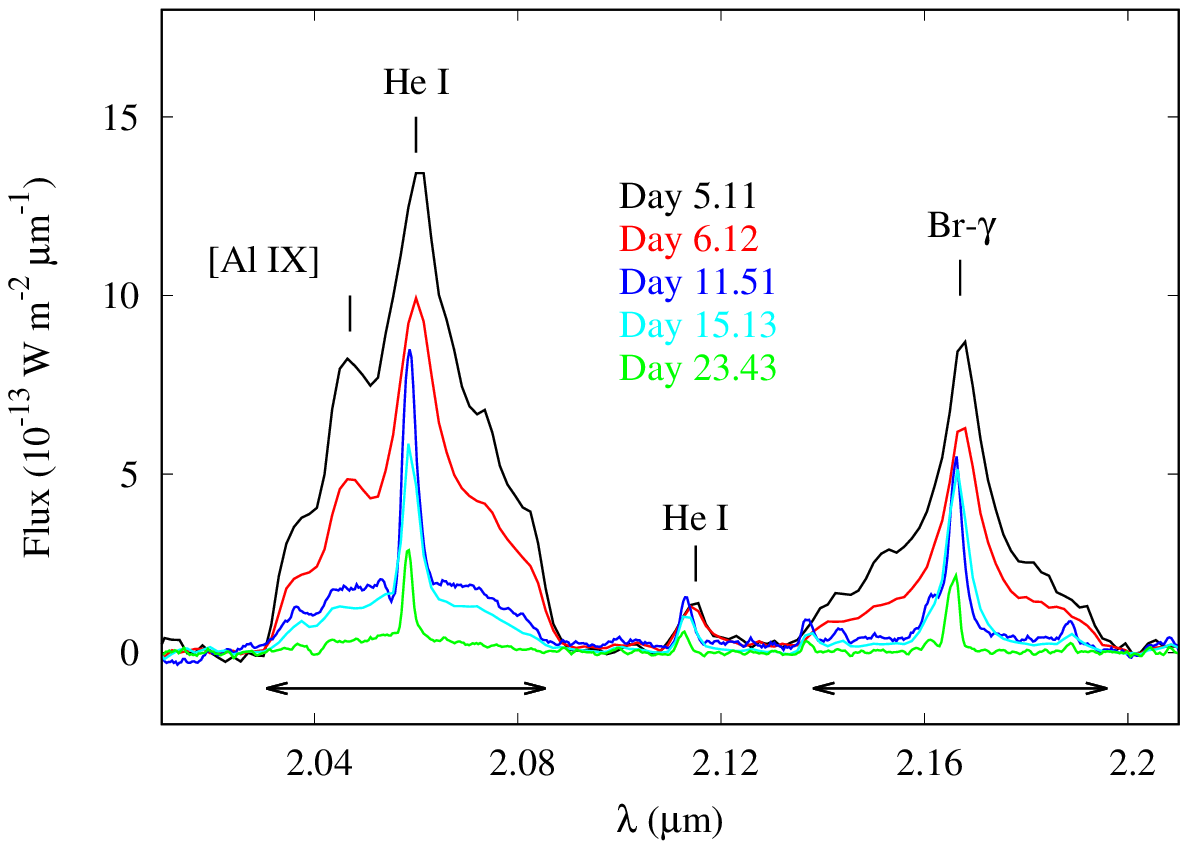}
 \includegraphics[width=7.cm]{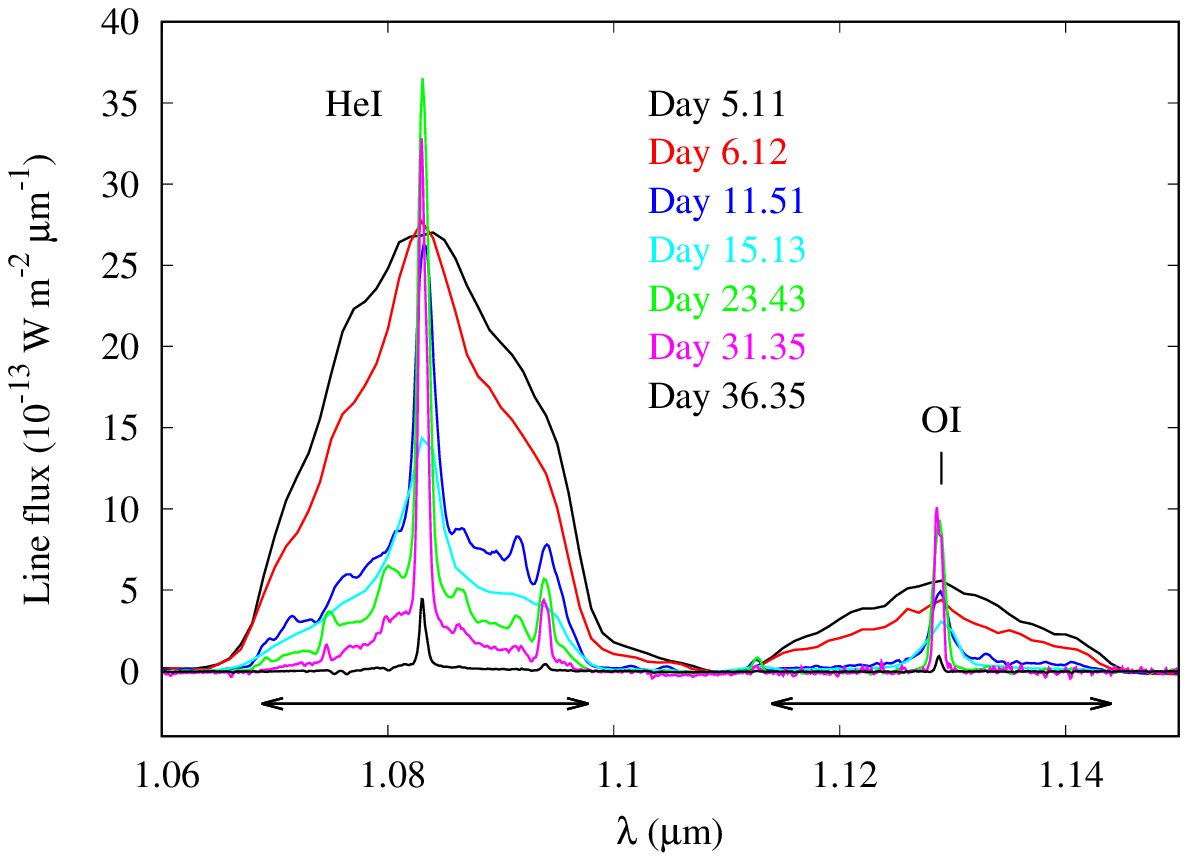}
 \includegraphics[width=7.cm]{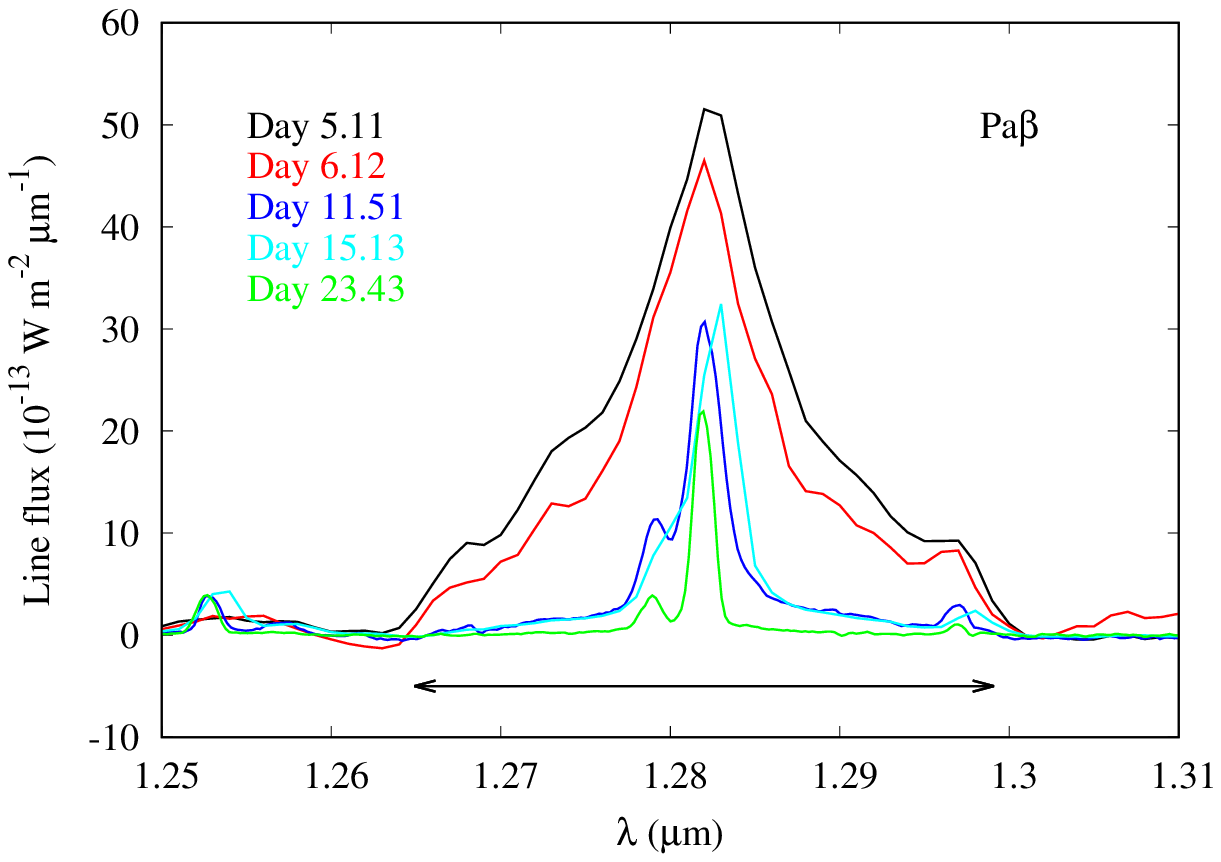}
\includegraphics[width=7.cm]{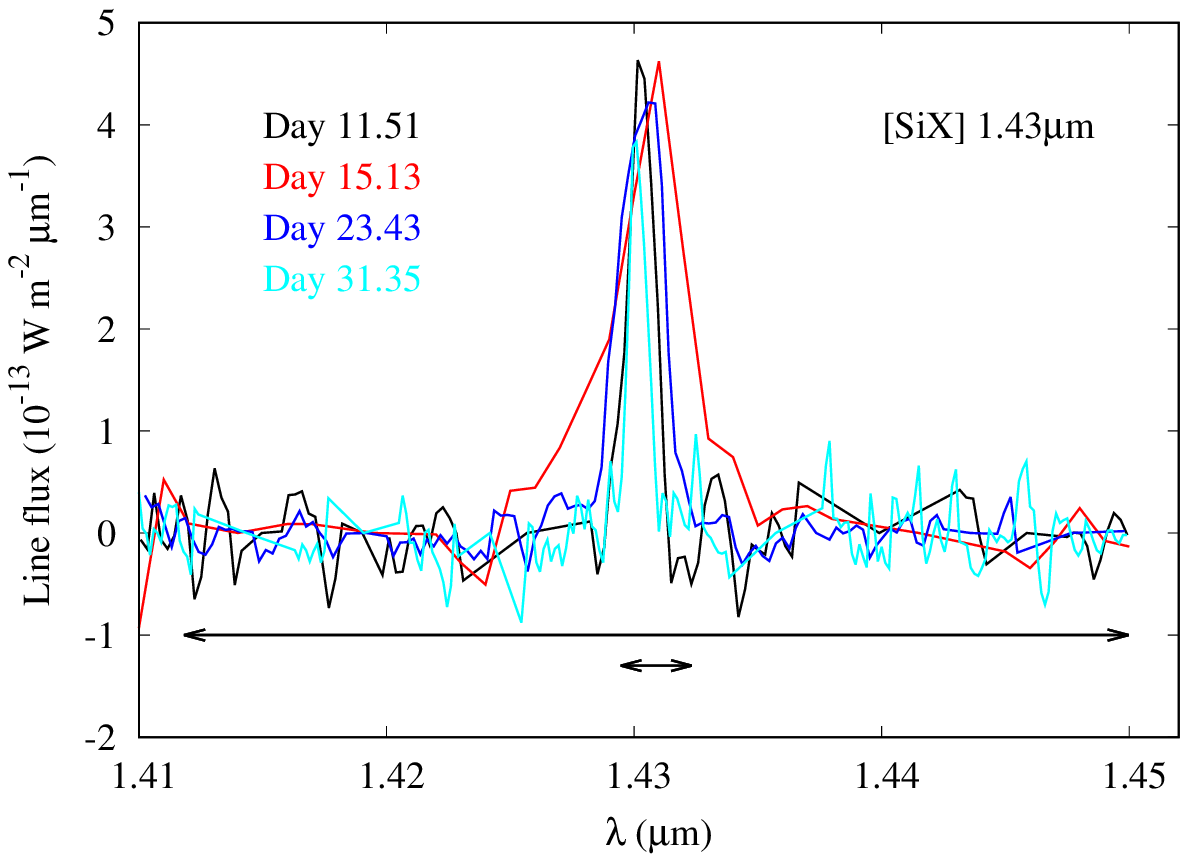}
\includegraphics[width=7.cm]{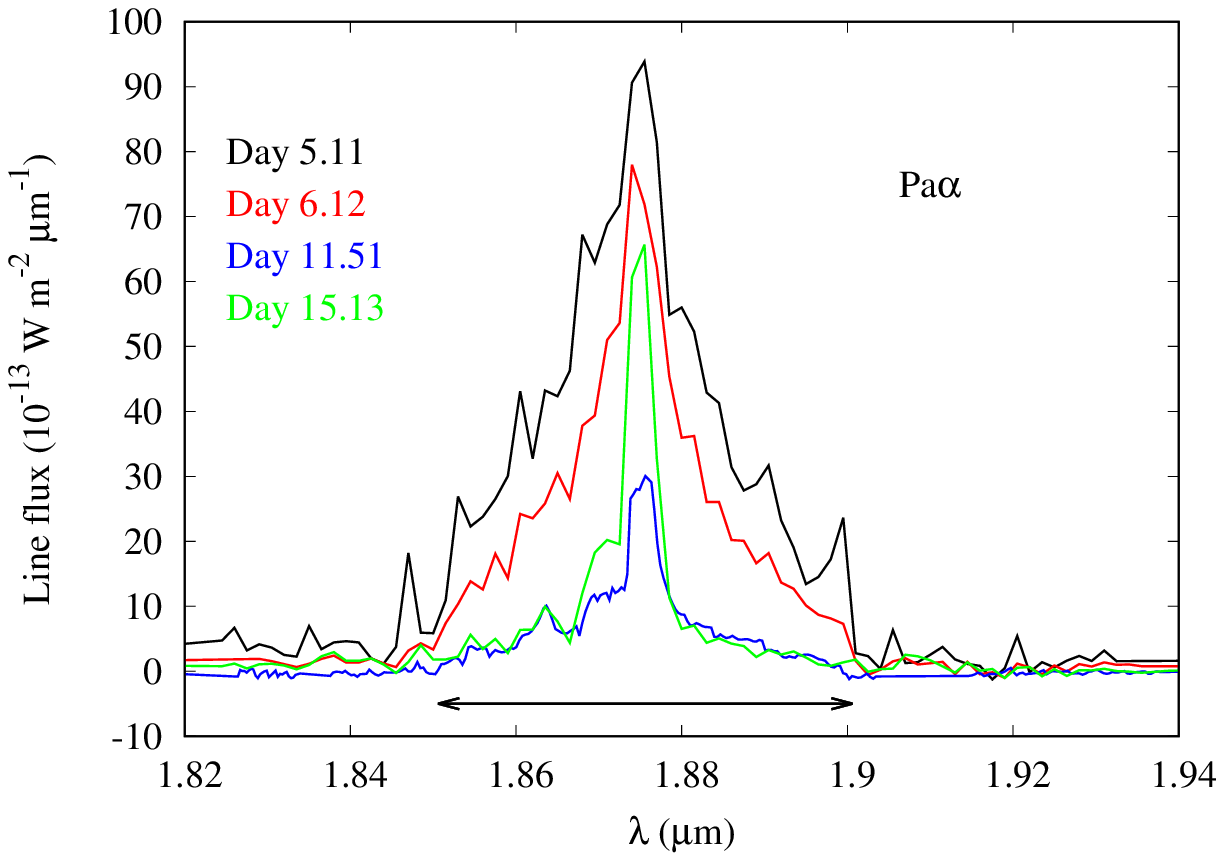}
\includegraphics[width=7.cm]{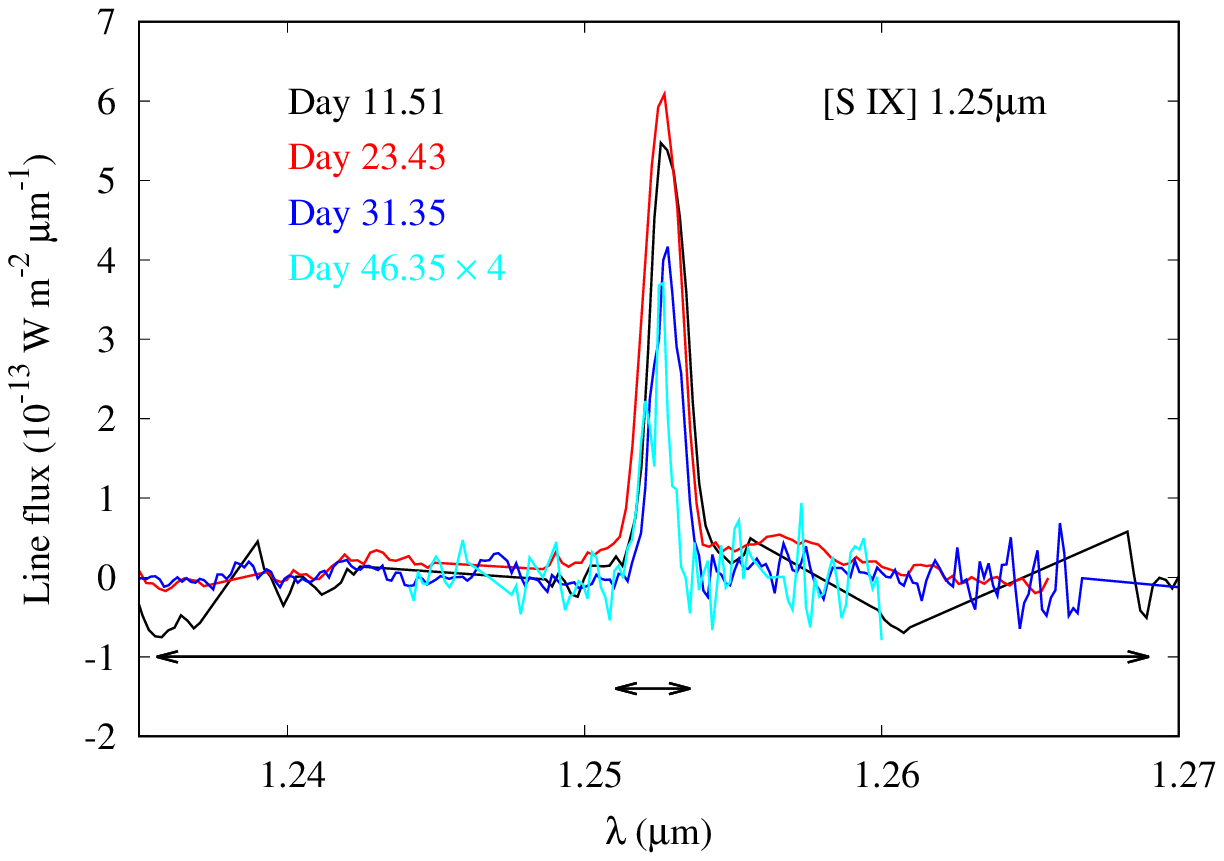}
\caption{Profiles of selected lines; in all cases 
the continuum has been subtracted.
Top left:  \pion{He}{i} 2.058\mic\ and Br\,$\gamma$.
Top right: \pion{He}{i} 1.083\mic\ and \pion{O}{i} 1.129\mic.
Middle left: Pa\,$\beta$; 
Middle right: \fion{Si}{x} 1.430\mic\ coronal line; the line is weak on the days
not shown. 
Bottom left: Pa\,$\alpha$.
Bottom right: \fion{S}{ix}.
Note the narrowness of the coronal lines.
In top, middle left and bottom left panels, the arrows depict $\pm4000$\vunit.
In the  middle and lower right panels the short arrows depict 
$\pm300$\vunit, the long arrows $\pm4000$\vunit. \label{2mic}}
\end{figure*}

\begin{figure}
\includegraphics[width=7cm]{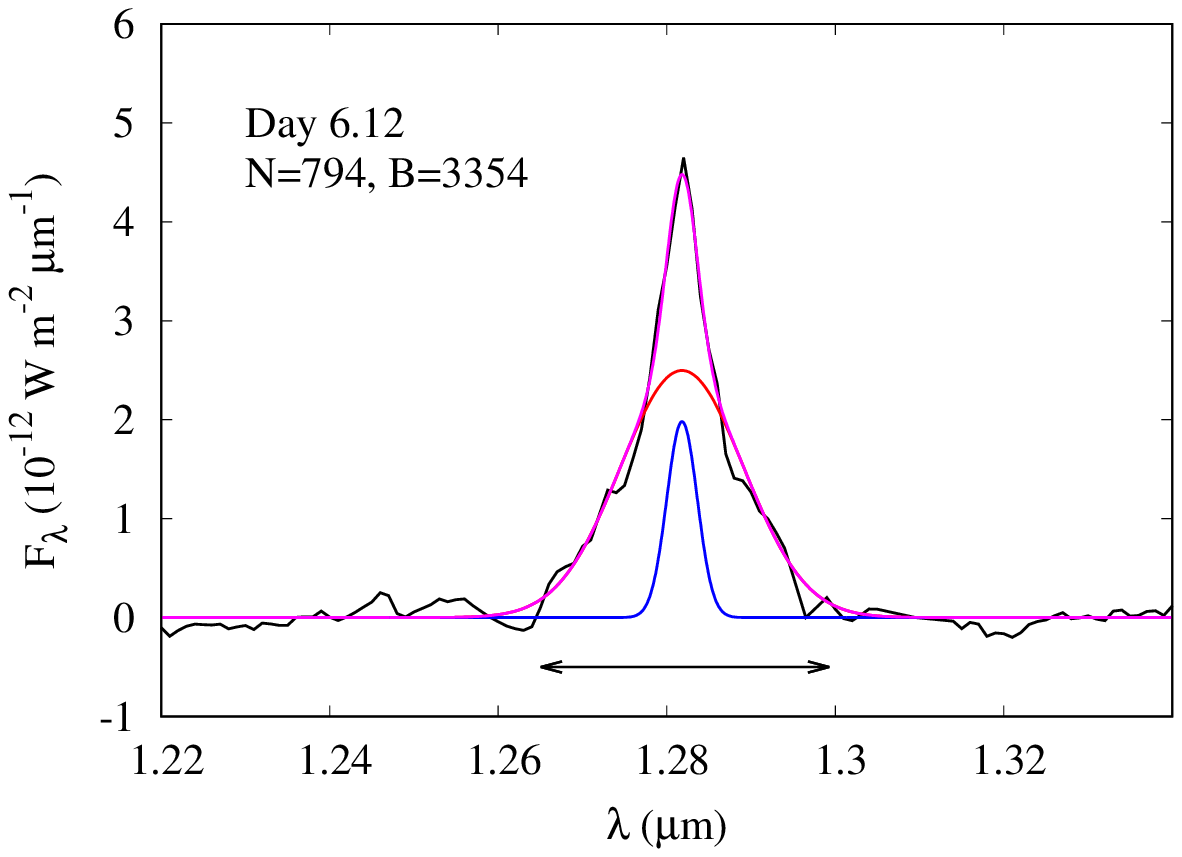}
  \includegraphics[width=7cm]{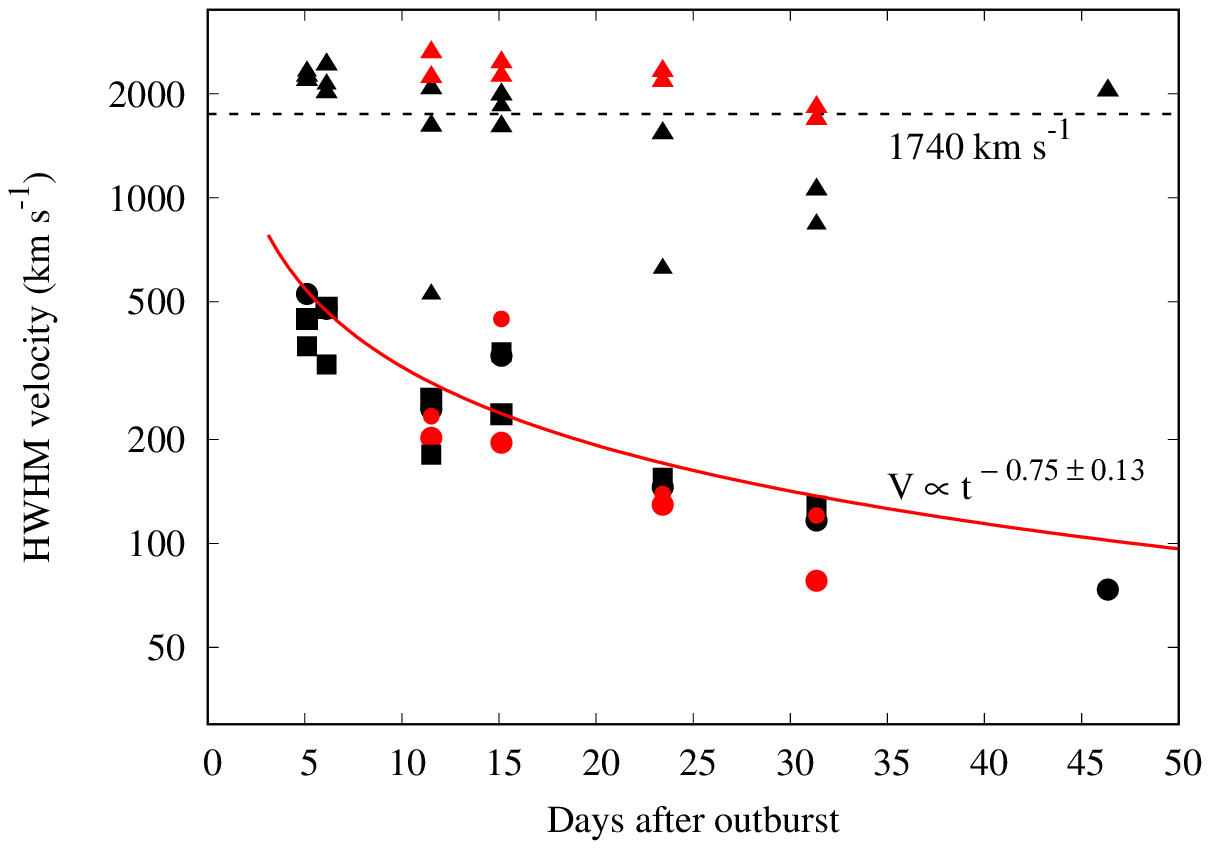}
  \includegraphics[width=7cm]{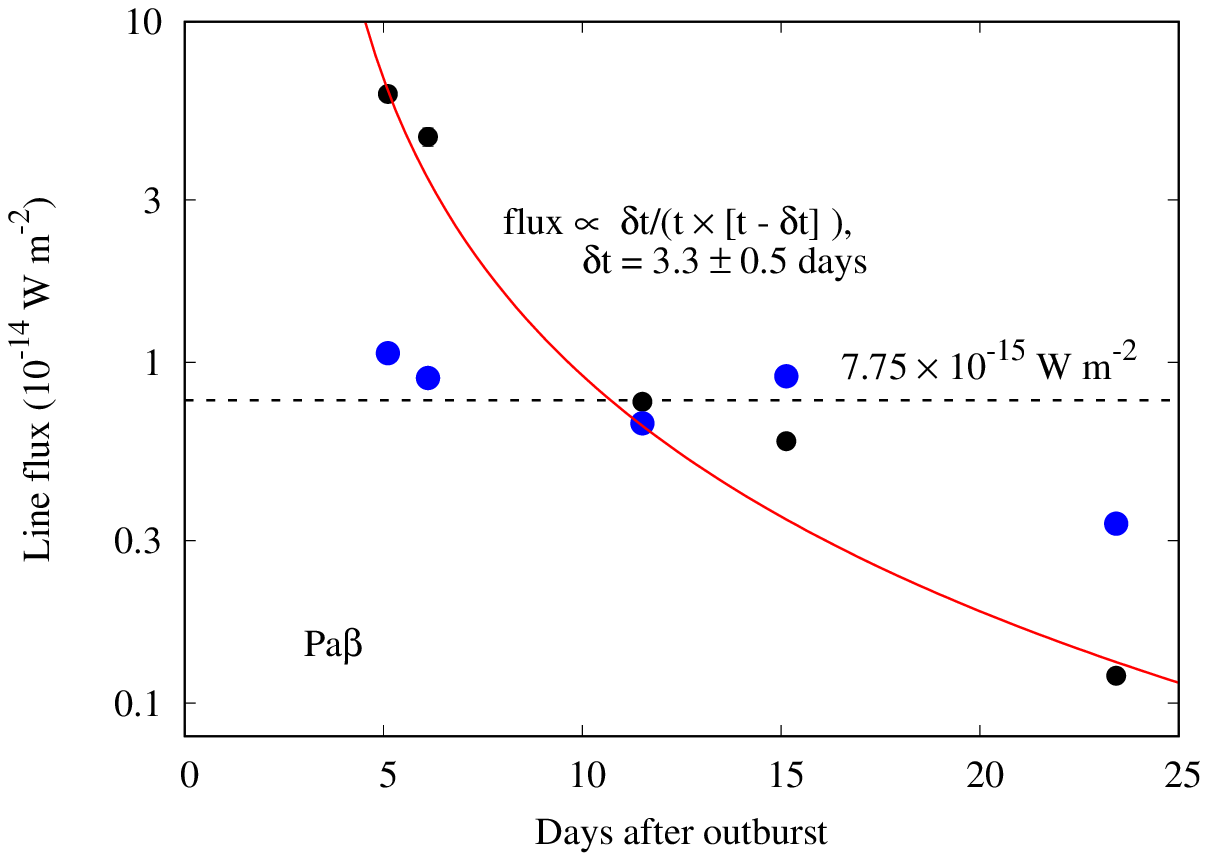}
 \caption{Top: fit of two gaussians to the continuum-subtracted 
 profile of Pa\,$\beta$ on day 6.12. Black curve: data; blue curve:
 narrow component; red curve: broad component; magenta curve: sum of the 
 two gaussians. The ``N'' and ``B'' values are the FWHM velocities, in
 \vunit, of the narrow and broad components, respectively, 
 after deconvolving the instrumental resolution.
 The arrows depict $\pm4000$\vunit.
 Middle: dependence of the mean HWHM velocities for the narrow 
  components of the Pa\,$\beta$ lines (black circles); the
  red curve is a fit of $V\propto{t}^{-\alpha}$ to the Pa\,$\beta$ data.
 Also shown are the data for the narrow components of Pa\,$\alpha$,
 Br\,$\gamma$ (black squares) and \pion{He}{i} (red circles).
 Broad components of \pion{H}{i} (black) and 
 \pion{He}{i} (red) are shown as triangles.
The broken horizontal line is the mean HWHM of the broad components
of the \pion{H}{i} lines.   
Bottom: dependence of observed line fluxes for Pa\,$\beta$ for
broad (black) and narrow (blue) components. The red curve is a fit of  
$f\propto{\delta{t}/[t(t-\delta{t})]}$ to the broad component.
The broken horizontal line is the mean flux of the narrow component.
 See text for details.
\label{H_decel}}
\end{figure}

The Pa\,$\alpha$, Pa\,$\beta$ and Br\,$\gamma$ lines were isolated and 
uncomplicated by the presence of other features throughout our 
observations. We fitted these lines using two gaussian components, 
one for the narrow core, the other for the broad component. 
For some dates it was also
possible to fit the \pion{He}{i} lines at 1.083\mic\ and 2.058\mic.
A typical fit is shown in the top panel of Fig.~\ref{H_decel}. 
We take the half width at half maximum (HWHM) of each component as
an estimate of the velocity of the bulk of the ejecta, after 
deconvolving the instrumental resolution (also assumed to be gaussian). 

The time-dependences of the HWHM velocities for the Pa\,$\alpha$, 
Pa\,$\beta$ and Br\,$\gamma$ and \pion{He}{i} lines, for both broad 
and narrow components, are shown in the middle panel of 
Fig.~\ref{H_decel}. There is a clear decline in the velocity 
widths of the narrow components, but little 
evidence of narrowing of the broad components. The expansion 
velocity as determined by the HWHMs of the narrow components of Pa\,$\beta$ 
(for which data exist for all seven dates) declines with time $t$, 
roughly as $1840[\pm490]\:\:t^{-0.75\pm0.13}$\vunit, with $t$ in days. 
Furthermore, the narrow components of the He lines follow the 
trend shown by the H lines very well. 

The behaviour of \rnsgr\ during its 
2019 outburst closely resembled that of other RNe and nova systems 
containing a giant secondary. Deceleration of the ejecta is a common
trait. This was seen in the eruptions of the RNe RS~Oph 
\citep{das06,evans07a,pandey22,woodward22} and V745~Sco \citep{banerjee14}, 
and in the symbiotic Nova Sco 2015 \citep{srivastava15}. 
It was also seen in the 2010 eruption of V407~Cyg, \citep[][their Fig.~3]{munari11,
banerjee14}, 
a system that has a Mira secondary displaying $\sim760$~d pulsations. Further, 
these systems \citep[with the exception of Nova Sco 2015, which is 
probably too distant;][]{srivastava15} were detected as $\gamma$-ray sources
during outburst.

On the other hand, the expansion velocity of the material giving
rise to the broad components seems relatively constant
(Fig.~\ref{H_decel}); the mean HWHM velocity of the 
broad component of the H lines is $1740\pm300$\vunit. The same
behaviour is shown by the broad components of the He lines.
For them the FWZIs of the broad components
are $\sim8000$\vunit, although for the He lines the FWZIs are less
certain because of blends with other lines.

The time-dependence of the line fluxes $f$, as typified by
Pa\,$\beta$, is shown in the bottom panel of Fig.~\ref{H_decel}.
The fluxes from the narrow components decline relatively little
over the course of our observations. 
The fluxes from the broad components, on the other
hand, decline by a factor $\gtsimeq30$ over a period
of $\ltsimeq25$~days, to such an extent that this component is hardly 
perceptible by day~46.35.

The general behaviour of the broad components may be understood in
terms of a simple model in which the material is ejected into a cone 
over a brief time interval $\delta{t}$, starting at $t=0$, 
and disperses thereafter at constant velocity; we refer to this below
as the ``polar component''. For a given 
emission line (such as Pa\,$\beta$), the only variables in 
this model are the density of the emitting material, $N_H$, and the 
electron density, $n_e$, both of which we take to vary with distance 
$r$ as $r^{-2}$. We assume that the electron temperature, mass-loss 
rate during ejection etc., remain constant. 
In these circumstances, 
the line flux $f$ follows a simple dependence on time:
\begin{eqnarray}
 f & \propto & \int_{R_1}^{R_2}  \:\: n_e \: N_H \:\: r^2 \:\: dr \nonumber \\ 
 & \propto & \frac{\delta{t}}{t (t - \delta{t}) } \:\: .
 \label{blob}
\end{eqnarray}
Here $R_2=Vt$ and $R_1=V(t-\delta{t})$ are the outer and inner radii
of the ejecta respectively.
A fit of this function to the data in the bottom panel of
Fig.~\ref{H_decel} shows a possible fit to the Pa\,$\beta$
data, with $\delta{t}=3.16\pm0.48$~days. A similar result is obtained
for Br\,$\gamma$ ($\delta{t}=3.37\pm0.41$~days). The fit for Pa\,$\alpha$, with
only four data points, is much poorer, $\delta{t}=1.71\pm1.28$~days.
The weighted mean for Pa\,$\beta$ and Br\,$\gamma$
gives $\delta{t}=3.3\pm0.5$~days.

The behaviours of the H recombination and \pion{He}{i} line profiles, 
together with those of a few other species, is further illustrated in 
Fig.~\ref{2mic}. The top left panel of this figure shows the 
evolution of the spectrum at 2.0--2.2\mic, which contains the 
\pion{He}{i} 2.058\mic\ and Br\,$\gamma$ lines. As the eruption 
progressed, the extrema of the broad components 
remained roughly at $\pm4000$\vunit, while the narrow
components became progressively narrower. 
The top right panel of 
this figure shows the evolution of the spectrum 
in the region of the \pion{He}{i} 1.083\mic\ and
\pion{O}{i} 1.129\mic\
lines; again the widths of the broad pedestals in the \pion{He}{i}
and \pion{O}{i} lines are constant.
The middle and bottom left panels show the evolution
of Pa\,$\beta$ and Pa\,$\alpha$ respectively; their behaviour is similar 
to that of Br\,$\gamma$.
The middle and bottom right panels show that the \fion{Si}{x} 1.43\mic\
and \fion{S}{ix} 1.25\mic\ coronal lines are narrow and essentially 
unresolved throughout, behaviour that is shared by all coronal lines.

\subsubsection{Coronal lines}
 \label{coronal_s}
\begin{figure*}
\includegraphics[width=14cm]{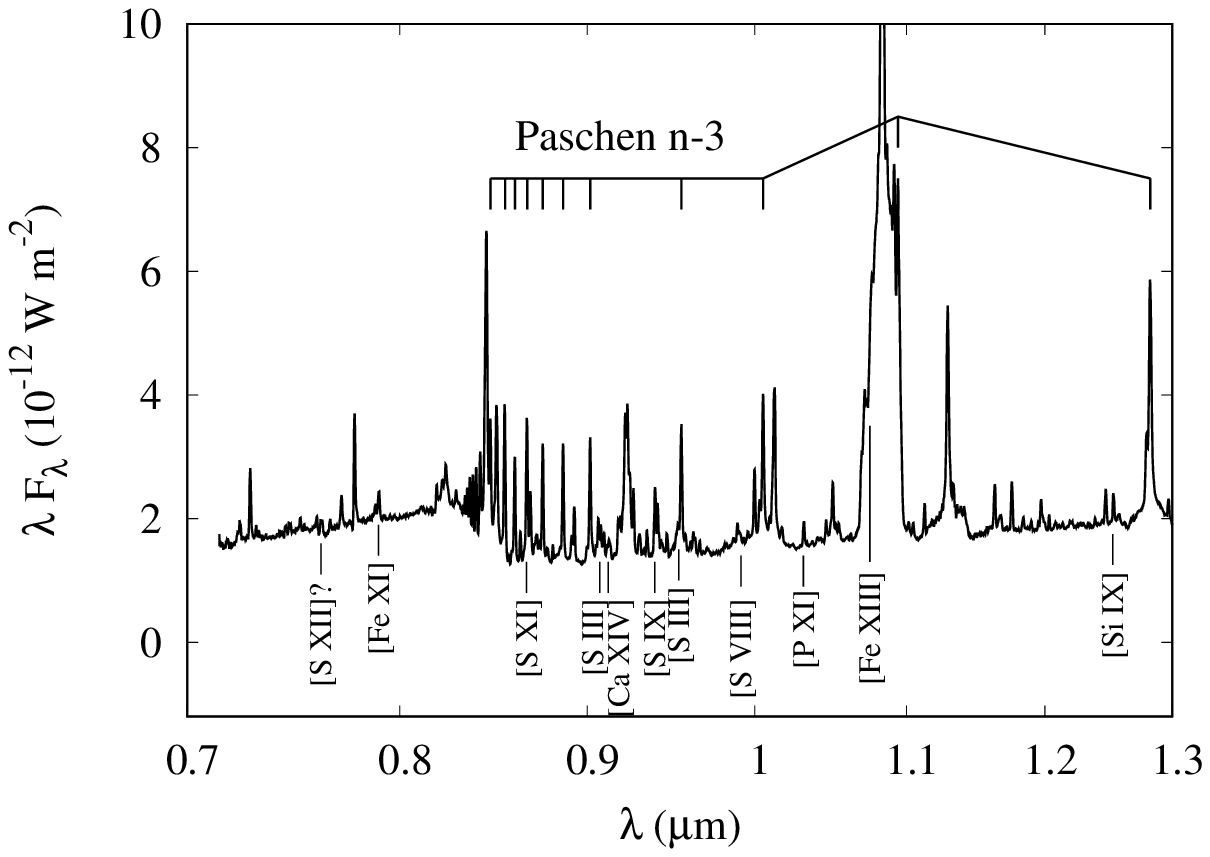}
\includegraphics[width=14cm]{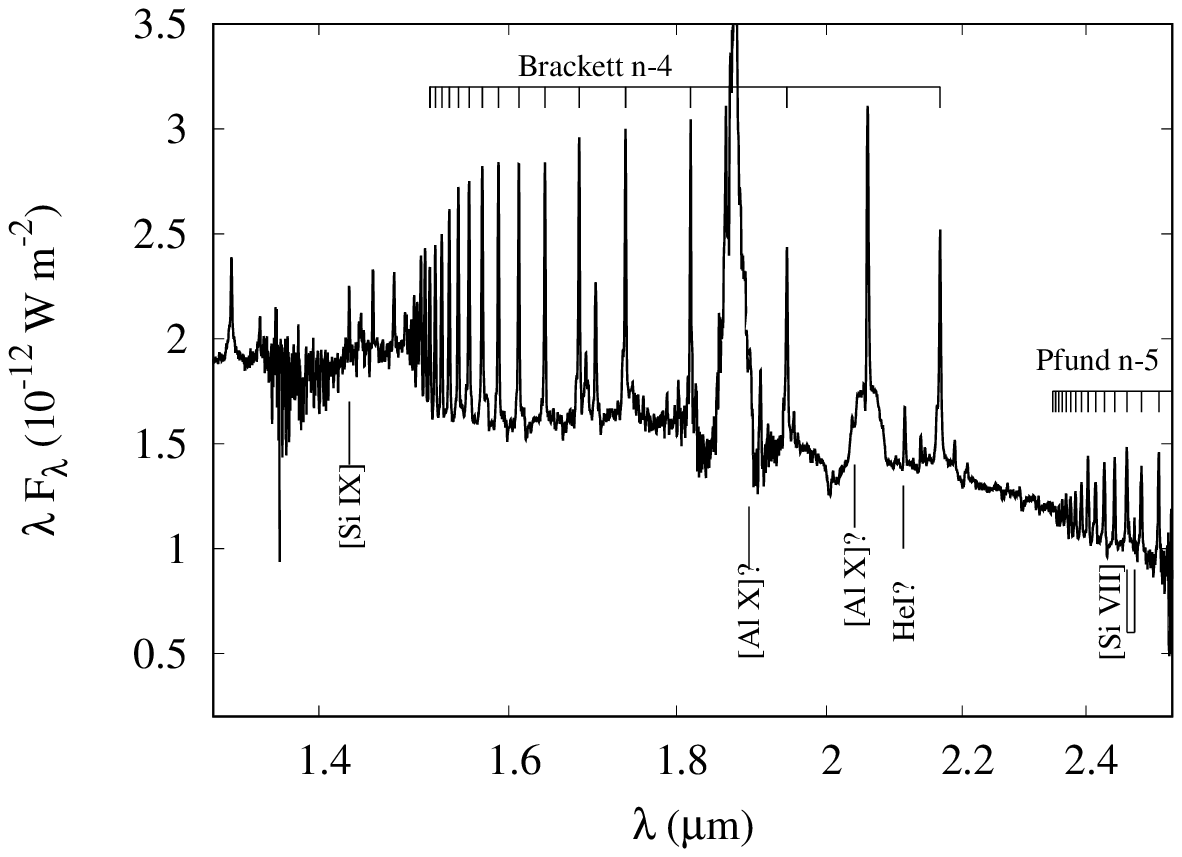}
\caption{Identification of \pion{H}{i} and coronal lines for 2019 September 8th (day~11.51).  
Top: wavelength range 0.7--1.3\mic, bottom: wavelength range 1.3--2.5\mic.
In both panels wavelength is plotted on logarithmic scales to 
reduce line-crowding at the shorter wavelengths.\label{all_1}}
\end{figure*}

\paragraph{Coronal line fluxes.} The IR spectra contain a number of coronal lines 
\citep[defined by][as emission lines ``arising from ground-state 
fine-structure transitions in species with ionisation potential 
$>100$~eV'']{greenhouse90}. 
These, together with their measured dereddened fluxes, are 
listed in Table~\ref{coronals}; a representative spectrum
(day~11.51) is shown in Fig.~\ref{all_1}, which identifies H recombination
and coronal lines. Many of these lines were present 
in the IR spectra of the 1985 and 2006 RN eruptions of RS~Oph \citep{evans88,evans07a,evans07b,banerjee09}.

The most likely identification of the feature at 2.111\mic\ 
(see Fig.~\ref{2mic}) is \pion{He}{i} ($^3$S$-^3$P), which 
is sometimes present in the NIR spectra of CNe 
\citep[see, e.g.,][]{naik09,raj12,raj15}. Its width is 
more comparable with those of other coronal lines
(compare its width with that of \pion{He}{i} 2.058\mic\ in
Fig.~\ref{2mic}) suggesting that it too might be a coronal line,
although it is weak so that any broad component might be lost in 
the noise.  An alternative identification might be 
\fion{Ca}{ix} ($^3$P$_2-^3$P$_0$) 2.111\mic.

Fluxes of isolated coronal lines were mostly determined 
by fitting gaussians to the line profiles; where a
coronal line is on the wing of a stronger
feature, or where the continuum is evidently non-linear, 
a quadratic polynomial was fitted to the continuum,
otherwise a linear continuum was used. 
In some cases the line sits on the wing of a much stronger line 
(e.g. \fion{Al}{ix} 2.045\mic\ on the blue wing of \pion{He}{i} 
2.058\mic; see Fig.~\ref{2mic}), making the line flux determination
somewhat less reliable.
In instances where the emission lines are crowded, the flux has been 
determined by trapezoidal integration between the FWZI points.

\paragraph{Temperature of the coronal gas.} 
The coronal line fluxes may be used to estimate the temperature of
the gas in which the coronal lines originate, using
\begin{equation}
  \frac{f(\mbox{A})}{f(\mbox{B})} = \frac{n\mbox{(A)}}{n\mbox{(B)}} \:
 \frac{\lambda_{\rm B}}{\lambda_{\rm A}} \:
 \frac{\Omega\mbox{(A)}}{\Omega\mbox{(B)}} \: \frac{g_{\rm B}}{g_{\rm A}} \:\:
 \label{green}
 \end{equation}
\citep{greenhouse90}; here A and B are two ionic states from the same 
atomic species, the $f$\,s are dereddened fluxes, the $n$\,s are the 
number densities of the ions, the $\Omega$\,s are effective collision 
strengths, and the $g$ values are the statistical weights of the lower 
levels. We take effective collision strengths from the IRON 
project\footnote{http://cdsweb.u-strasbg.fr/tipbase/home.html}
\citep{hummer93,badnell06}. 
All ionisation fractions are from \cite{arnaud85}, other than
P\,{\sc xi}, which is not listed in \citeauthor{arnaud85}; we take the 
value for $T=10^6$~K from \cite{jain78} for P\,{\sc xi}.

We assume an initial electron temperature $T$, and calculate the 
effective collision strengths (which are $T$-dependent); this gives
a value for $n$(A)/$n$(B). This in turn gives a new value of $T$. 
The process is iterated until convergence occurs, which it does rather 
quickly. Where the Iron Project data do not extend to suffiently high
temperatures (usually above $10^5$~K), 
we use the highest temperature available in the database
(see Table~\ref{coronals} for these cases).
Collision strengths and ionisation fractions are interpolated from
tabulated values by fitting polynomials to the latter.

For two species we have spectral lines from three ionic states, 
namely Si (\fion{Si}{vi}, \fion{Si}{vii} and
\fion{Si}{x}) and S (\fion{S}{viii}, \fion{S}{ix} -- two transitions -- and
\fion{S}{xii}). We use the dereddended line fluxes for these species 
for day 23.43, when (a)~there has been substantial deceleration of the 
ejecta (see Fig.~\ref{H_decel}) and (b)~we expect photoionisation to be 
small towards the end of the SSS phase. Using Equation~(\ref{green}) 
for Si, we obtain $\log{T}$ values  ($T$ in~K) of
5.69 (\fion{Si}{vi} and \fion{Si}{vii}) and
5.96 (\fion{Si}{vii} and \fion{Si}{x}). For S we get
5.95 (\fion{S}{viii} and \fion{S}{ix})
6.17 (\fion{S}{ix} and \fion{S}{xii}) and
6.09 (\fion{S}{viii} and \fion{S}{xii}).
Using these two species we find a mean $\log{T}= 5.97\pm0.11$, or 
$T=9.3^{+2.7}_{-2.1}\times10^{5}$~K on day 23.43. 
We assume $T=9.3\times10^5$~K ($kT=80$~eV) in what follows.
The final values of $\Omega$ used are listed in Table~\ref{coronals}.

\paragraph{Abundances in the coronal gas.} 
Further, by applying Equation~(\ref{green}), now with A and B representing
different species \citep[see][]{greenhouse90}, we can determine 
the relative abundances of ionic species and, using the ionisation 
fraction as a function of temperature \citep{arnaud85}, 
the abundances of the elements themselves.

We have determined the abundances of each species X, by number, 
relative to Si,
using ion pairs in the ``matrix'' of ions in Table~\ref{abund}. 
The abundance ratios as determined relative to \fion{Si}{vii} and
\fion{Si}{x} are satisfyingly consistent for each ion --- although there
are differences between ions -- while those determined using
\fion{Si}{vi} are consistently $\sim1.4$~dex lower. For our present purposes,
we take an overall mean, using all values for each species.
These averaged abundance ratios, by number relative to Si, are given in
the bottom row of Table~\ref{abund}.

\begin{figure}
\centering
 \includegraphics[width=8cm]{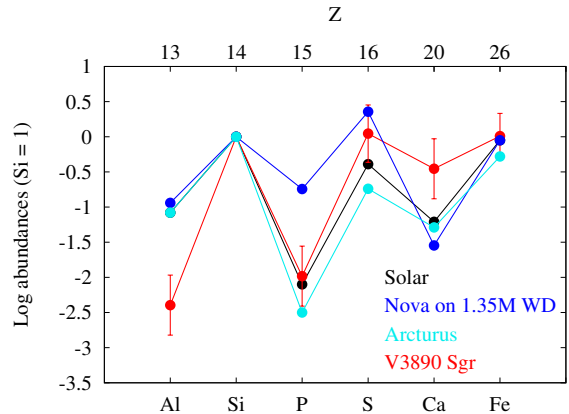}
 \caption{Abundances in the ``equatorial'' component of \rnsgr,
 i.e., in the orbital plane.
 Black points and lines: solar photospheric abundances from
 \protect\cite{asplund21}. 
 Blue points and lines: abundances predicted for a TNR on the surface
 of a 1.35\Msun\ CO WD \citep{starrfield20}; 
 the abundances shown
 are for a 50:50 mix of WD and accreted material.
 Cyan points and lines: abundances for the K2 giant Arcturus 
 \citep{peterson93}.
 Red lines: abundances in \rnsgr.
  See text for details.
 \label{abundf}}
\end{figure}

The abundances are shown in Fig.~\ref{abundf}, together with the 
corresponding solar \citep*{asplund21} and Arcturus 
\citep[K2III;][]{peterson93} values, for comparison. 
\cite{pavlenko08, pavlenko20} 
and \cite{kaminsky22} have shown that the abundances of some of the 
light elements (specifically C, N, O, Si) differ from solar by less 
than one dex for the RG components of RNe, so the solar values provide
a reasonable template. Also included in 
Fig.~\ref{abundf} are the expected abundance ratios for a TNR on the 
surface of a 1.35\Msun\ WD, with 50:50 mixing of WD and accreted 
material \citep{starrfield20}; solar abundance has been assumed for Fe 
on the premise that Fe is not produced in a nova TNR.

Considering the crude nature of our abundance determination, the abundance 
ratios for \rnsgr\ are, with the exception of a deficiency of Al, 
surprisingly closer to the solar and Arcturus values than they are 
to the predictions of TNRs. However we should be mindful of the fact that
\rnsgr, at $D=9$~kpc and close to the direction of the Galactic Centre
(Galactic longitude $9\fdg2$), is much closer to the Galactic Centre 
than either the Sun or Arcturus, and abundances in its RG component might 
also reflect metallicity gradients in the Milky Way 
\citep[see, e.g..][]{maciel10}.

\begin{landscape}
 
 \begin{figure*}
 \includegraphics[width=5.4cm]{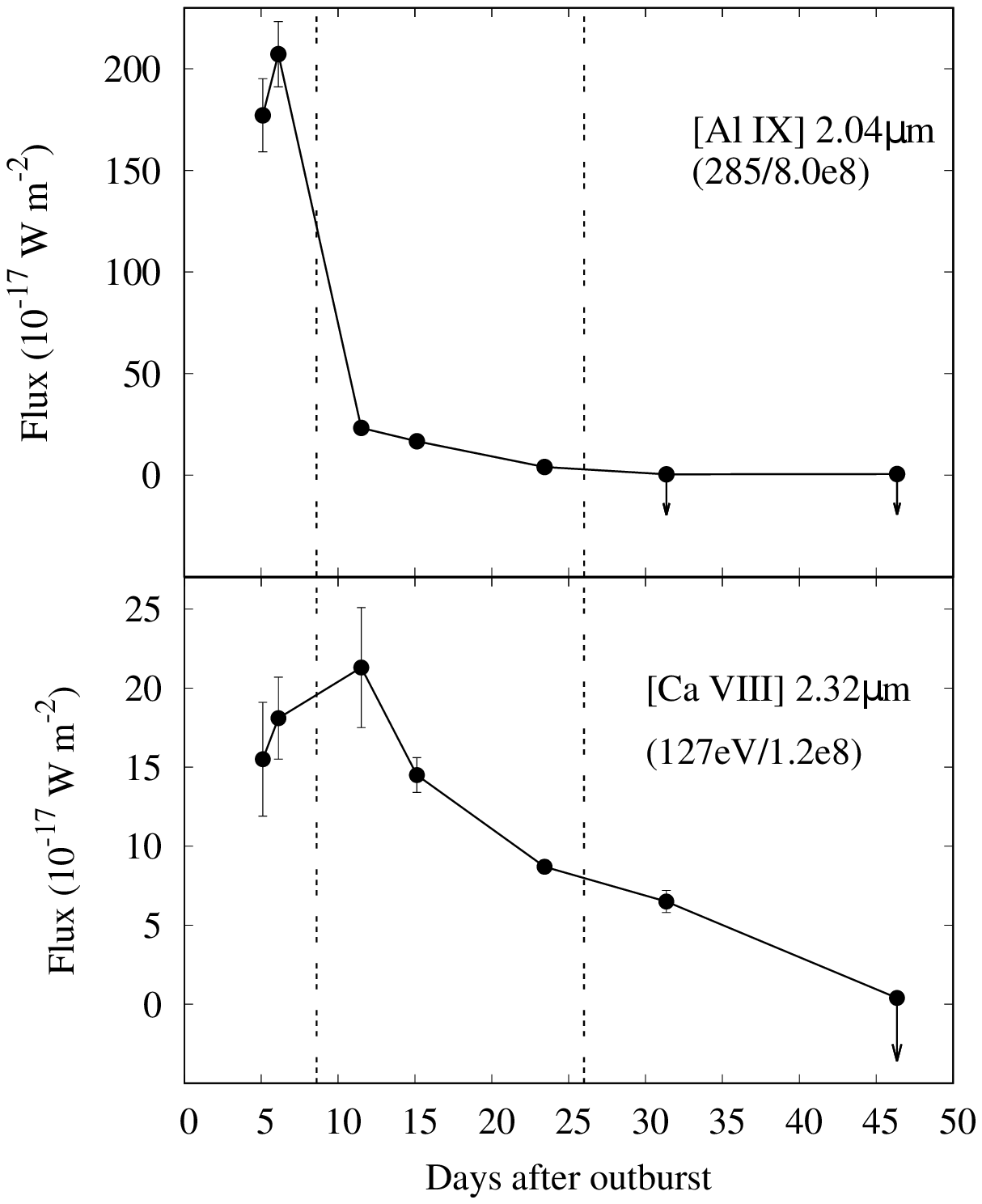}
\includegraphics[width=5.7cm]{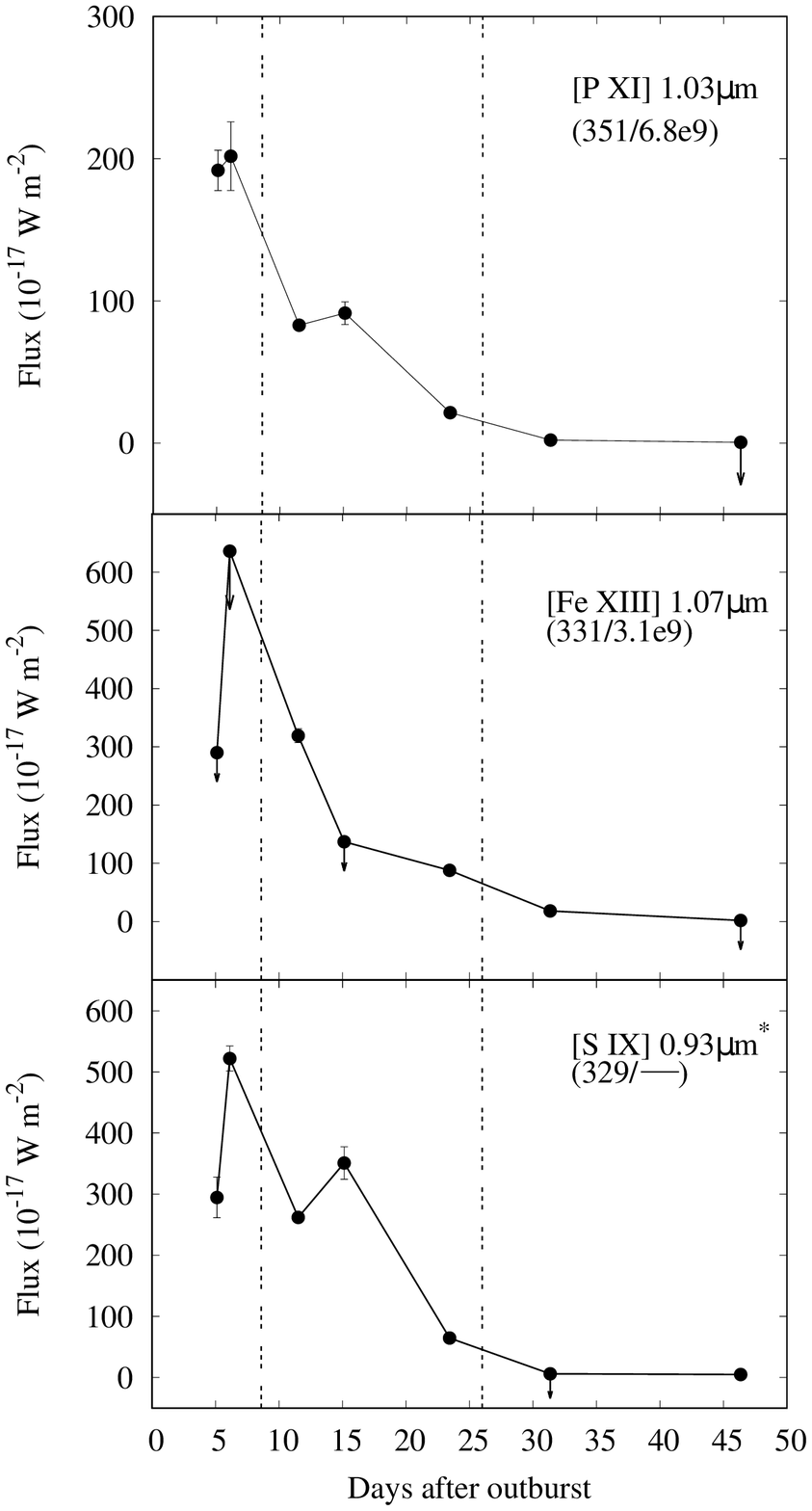}
 \includegraphics[width=5.6cm]{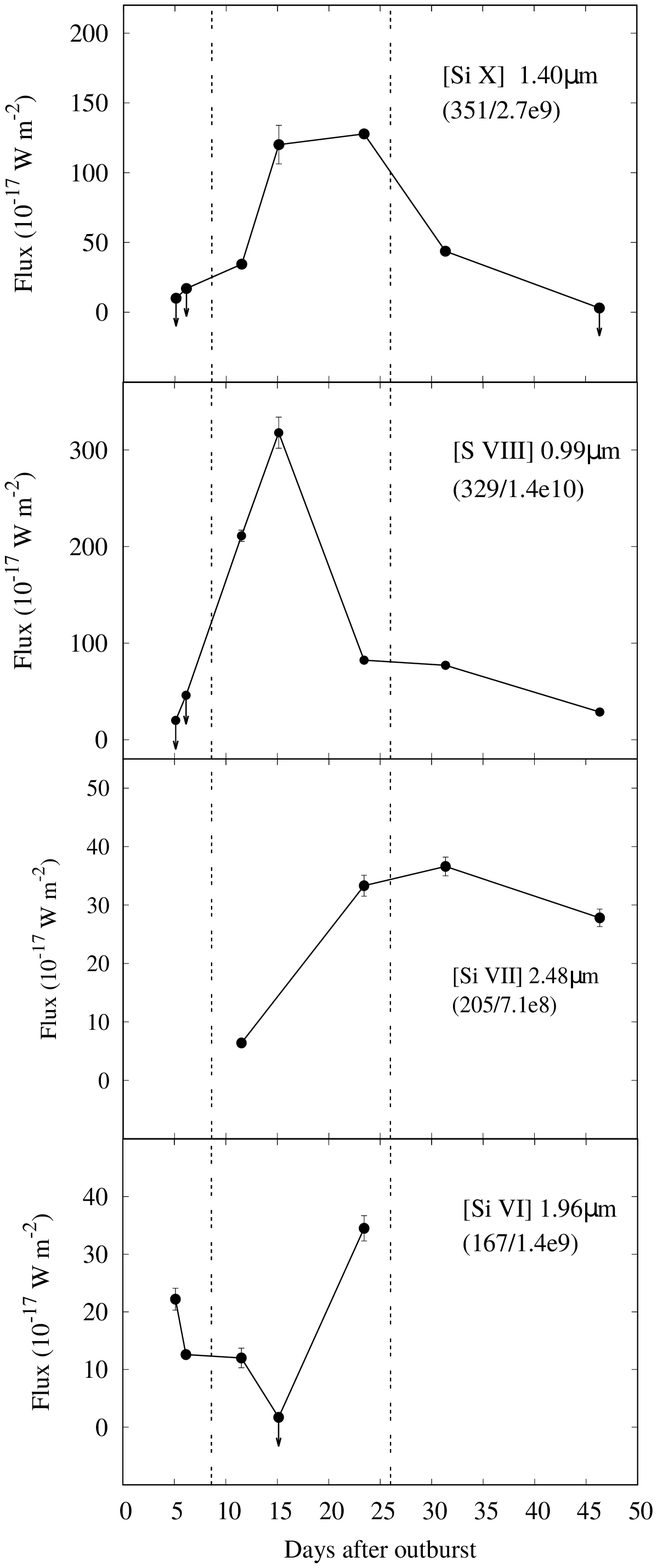}
\caption{Time-dependence of the dereddened aluminium, calcium,
iron, phosphorus, silicon and sulphur coronal line fluxes from the 
2019 eruption.
Left and centre: coronal line the fluxes of which peak early in the 
eruption.
Right: coronal line the fluxes of which peak later in the eruption.
For lines marked with an asterisk ``*'', some of the features are in a crowded  
region of the spectrum and fluxes may not be completely reliable.
Numbers in brackets below identification denote IP in eV/critical 
electron density (in cm$^{-3}$) from Table~\ref{critical}, in the form 
XeY = $X\times10^Y$. See text for details.
Vertical dotted lines denote approximate start and end of the SSS phase, from \protect\cite{page20}.
\label{coronal_lc}}
\end{figure*}
 
 \end{landscape}

\paragraph{Coronal line ``light curves''.}
The time-dependences of the dereddened fluxes for some selected coronal
lines (silicon, sulphur, calcium and aluminium) are shown in 
Fig.~\ref{coronal_lc}. There is a hint that some lines peak early 
in the eruption (e.g. \fion{Al}{IX} 2.04\mic),
while others (e.g. \fion{Si}{vii} 2.48\mic, \fion{Si}{x})
peak later (\gtsimeq20~days).

\begin{landscape}

\begin{table*}
\caption{Coronal line fluxes. Fluxes, dereddened by $E(B-V)=0.5$, 
are in $10^{-17}$~W~m$^{-2}$;  the flux for \fion{P}{viii} (1.2784\mic)
includes contributions from both \fion{P}{viii} and \fion{Cr}{ix}.
Upper limits are $3\sigma$.
Wavelengths, from van Hoof (2018), are vacuum values.
Ionisation potentials of the lower
ionisation stage (IP; from the NIST Database$^\dag$) are in eV.
Effective collision strengths $\Omega$ are at $\log{T}\mbox{(K)}=5.97$ unless otherwise specified; see text for details.
Fluxes marked ``*'' denote lines in a crowded region of the spectrum, or lines
that lie on the wing of a stronger line, and fluxes may not be completely reliable.
A ``---'' denotes that the line was ouside the observed
wavelength range. 
\label{coronals}}
\begin{tabular}{cccccccccccc}
  \multicolumn{3}{c}{Line identification}&  \multicolumn{7}{c}{Day} \\ \cline{1-3}\cline{6-12} \noalign{\smallskip}
ID & $\lambda$ ($\mu$m) & Transition $u-\ell$      &IP &  $\Omega$  & 5.11   &   6.12   &   11.51     &   15.13      &   23.43       & 31.35         &   46.35   \\\hline\hline
\fion{S}{xii} & 0.7613 & $^2$P$_{3/2}-^2$P$_{1/2}$ &505&  0.465 & ---      &  ---     &$95.8\pm5.8$ & ---          & $284.2\pm3.0$ & ---           &---  \\    
\fion{Fe}{xi} & 0.7894 & $^2$P$_{3/2}-^2$P$_{1/2}$ &262&  0.887& ---      &  ---     &$157.4\pm5.3$ & ---          & $408.6\pm3.7$ & ---           &  ---   \\
\fion{S}{ix} & 0.9391 & $^3$P$_{0}-^3$P$_{2}$      &329& $0.264^\ddag$ & $294.6\pm33.2^*$  
                                                                   & $522\pm20.5^*$ &$261.9\pm5.1$&$351.0\pm26.6$&$64.6\pm2.8$   &  $<6$   &$4.9\pm0.8^*$\\
\fion{S}{viii} & 0.9914 & $^2$P$_{1/2}-^2$P$_{3/2}$&281& 0.298 & $<20$      &$<46$     &$211.1\pm5.8$& $317.8\pm16.1$    &$82.4\pm1.1$   &$77.0\pm 0.7$  &$28.7\pm0.6$ \\
\fion{P}{xi} & 1.0310 & $^2$P$_{3/2}-^2$P$_{1/2}$  &424& $0.543^\ddag$  & $191.8\pm14.2^*$
                                                                   &$201.8\pm24.1$
                                                                              &$82.9\pm1.6$ &$91.4\pm8.0$  &$21.4\pm0.6$   & $2.1\pm0.2^*$  & $<0.6$  \\
\fion{Fe}{xiii} & 1.0750 & $^3$P$_{2}-^3$P$_{1}$   &331&  2.638  &$<290$     &  $<636$  &$319.1\pm11.9$&  $<137$  &$88.0\pm4.6$   & $18.2\pm1.0^*$  & $<1.9$ \\ 
\fion{S}{ix} & 1.2523 & $^3$P$_{1}-^3$P$_{2}$      &329& $0.264^\ddag$ & $370.9\pm46.1^*$  &  $350.5\pm21.9^*$ &$83.3\pm2.8$ & $202.2\pm15.6$  &$94.6\pm1.1$ & $46.2\pm0.7$ & $7.9\pm0.8$   \\
\fion{P}{viii} & 1.2784 & $^2$P$_{0}-^2$P$_{2}$    &264& --- & $<19$   &  $<47$    &$142.8\pm4.6$&  $<5.8$   &$80.9\pm3.4^*$   & $19.9\pm1.7^*$  & $6.5\pm0.9$    \\
 \fion{Cr}{ix} & 1.2786 & $^3$P$_{2}-^3$P$_{1}$    &185& --- &      &     &      &    &        &    &  \\
\fion{Si}{x} & 1.4309 & $^2$P$_{3/2}-^2$P$_{1/2}$  &351& 0.525 & $<10$   &  $<17$ &$34.4\pm2.2$ &$120.1\pm13.8$   &$127.8\pm2.8$  & $43.7\pm2.3$ & $<2$  \\
\fion{Si}{vi} & 1.9650 & $^2$P$_{1/2}-^2$P$_{3/2}$ &167& 0.380 & $22.2\pm1.9^*$ &  $12.6\pm0.7^*$  & $12.0\pm1.7^*$    &  $<1.7$   &$34.5\pm2.2$& ---       & --- \\
\fion{Al}{ix} & 2.0450$^*$  & $^2$P$_{3/2}-^2$P$_{1/2}$ 
                                                   &285& 0.604 & $177.1\pm18.0$ & $207.2\pm16.1$ & $23.3\pm1.4^*$ & $16.7\pm1.3^*$ & $4.0\pm0.6^*$ & $<0.4$ & $<0.6$ \\
\fion{Ca}{viii} & 2.321 & $^2$P$^o_{3/2}-^2$P$^o_{1/2}$ 
                                                   &127& 2.577 & $15.5\pm3.6^*$ & $18.1\pm2.6$ & $21.3\pm3.8$ & $14.5\pm1.1$ & $8.7\pm0.4$ & $6.5\pm0.7$ & $<0.4$ \\
\fion{Si}{vii} & 2.4833 & $^3$P$_{1}-^3$P$_{2}$    &205&  0.695 & ---  & ---   &$6.4\pm0.6$  &  ---    &$33.3\pm1.8$   & $36.6\pm1.6$    & $27.8\pm1.5$ \\\hline\hline
\multicolumn{10}{l}{$^\dag$https://www.nist.gov/.} \\
\multicolumn{10}{l}{$^\ddag$Value at $\log{T}\mbox{(K)}=5$.}
\end{tabular} 
 \end{table*}

 \end{landscape}

\begin{table*}
 \caption{Atomic abundances of X relative to Si, $\log$[$n$(X)/$n$(Si)], 
 as determined from coronal lines. See text for details. \label{abund}}
\begin{tabular}{cccccccccc} \hline\hline
 & \fion{Al}{ix} & \fion{P}{xi} & \fion{S}{viii} & \fion{S}{xi} & \fion{S}{xi} & 
 \fion{S}{xii} & \fion{Ca}{viii} & \fion{Fe}{xi} & \fion{Fe}{xiii} \\
 $\lambda(\mic)=$ & 2.043 & 1.031 & 0.9914 & 0.9391 & 1.2523 & 0.7613 & 2.321 & 0.7894 & 1.0750 \\  \hline
\fion{Si}{vi} & --3.246 & --2.834 & --1.423 & --1.397 & --1.688 & 1.285 & --1.305 & 
--1.272 & --0.412 \\
\fion{Si}{vii} & --1.914 & --1.502 & --0.091 & --0.065 & --0.356 & 2.617 &  0.027 & 
0.060 & 0.920  \\
\fion{Si}{x} & --2.025 & --1.613 & --0.203 & --0.177 & --0.468 & 2.506 & --0.085 &
--0.051 & 0.809 \\  \hline
&&&&&&&&&\\
 Adopted    & Al & P & \multicolumn{4}{c}{S} & Ca & \multicolumn{2}{c}{Fe} \\ \cline{2-10} \noalign{\smallskip}
abundances & $-2.40\pm0.43$ & $-1.98\pm0.43$ & \multicolumn{4}{c}{$0.05\pm0.41$} & $-0.46\pm0.43$& 
\multicolumn{2}{c}{$0.01\pm0.32$} \\
\hline\hline
  \end{tabular}
\end{table*}

While we have no NIR data before day~5.11, spectra in the region of 
H\,$\alpha$ in the ARAS 
database\footnote{http://www.astrosurf.com/aras/Aras\_DataBase/DataBase.htm}
\citep{teyssier19} for $t\ltsimeq5$~days indicate that its narrow 
component was becoming narrower some time before our first NIR 
observation, possibly from outburst: clearly the ejecta were
encountering significant resistance from near the time of outburst.
\cite{page19a} reported that the SSS phase of the 2019 eruption
started on day~9.12, so that the gas would have been ionised by the 
early X-ray emission. The early coronal emission may therefore have
arisen in a mix of shock-ionised and photoionised gas.
The subsequent decline in the coronal line fluxes, with an 
$e$-folding time of $\sim15$~days (cf. Fig.~\ref{coronal_lc}), 
may correspond to the decline of the SSS phase; 
\cite{page19b,page20} report that the SSS phase in the 2019 eruption 
of \rnsgr\ had effectively ended by day~26.
Fig.~\ref{coronal_lc} (left and centre panels) 
suggests that the early-peaking coronal lines had also
disappeared on this time-scale, suggesting that the coronal 
lines at the earlier times arose primarily in a photoionised gas,
with shock ionisation also playing a role.

However, regarding the later ($\gtsimeq23$~days) coronal emission,
it is likely that the lines arise in a gas that is solely 
collisionally ionised and excited. As the ejecta encounter the RG wind, 
decelerate and are shocked, we 
expect a resurgence in the coronal line emission. We see in 
Fig.~\ref{coronal_lc} that this results in the onset of emission 
in (for example) \fion{Si}{x} 1.430\mic, and the recovery of 
(for example) \fion{Si}{vi} 1.96\mic. It seems that the 
shock-origin for the coronal line emission sets in after 
$\sim15-20$~days, just as the SSS-related coronal lines begin
to fade.

\begin{table}
\caption{Critical electron density, determined as described in
text, for coronal lines in Fig.~\ref{coronal_lc}. \label{critical}}
\begin{tabular}{cccc}
 ID & $\lambda$ ($\mu$m)  & $A$ (s$^{-1}$)   & $n_{\rm crit}$ (in cm$^{-3}$) \\
    &                     &                  &  at  ${T}=9.3\times10^5$~K  \\\hline\hline
\fion{S}{viii} & 0.9914 & 18.64 & $1.4\times10^{10}$ \\
\fion{P}{xi} & 1.0310 & 8.19 & $6.8\times10^{9}$ \\
\fion{Fe}{xiii} & 1.0750 & 14.48 & $3.1\times10^{9}$ \\ 
\fion{Si}{x} & 1.4309 & 3.15 & $2.7\times10^{9}$ \\
\fion{Si}{vi} & 1.9650  & 2.38 & $1.4\times10^{9}$ \\
\fion{Al}{ix} & 2.0450 & 1.07 & $8.0\times10^{8}$ \\
\fion{Ca}{viii} & 2.321 & 0.70 & $1.2\times10^{8}$\\\hline\hline
\end{tabular} 
 \end{table}

There seems to be no obvious correlation of this behaviour with 
the IP values in Table~\ref{coronals}.
There is also no clear correlation with the critical
electron density, $n_{\rm crit}$, above which the upper level is collisionally, 
rather than radiatively, de-excited. The critical density 
at temperature $9.3\times10^5$~K is \citep[see, e.g.,][]{agn}
\[ n_{\rm crit} = 1.120\times10^{8} \:\: \frac{A\:g_2}{\Omega(T)}\] 
where $A$ is the Einstein coefficient for the transition, and $g_2$
is the statistical weight of the upper level. Einstein coefficients,
where available, have been taken from \cite{vanhoof18}. The resulting values of 
$n_{\rm crit}$ are given in Table~\ref{critical}.
The high values of $n_{\rm crit}$ are a consequence of the high
temperature of the coronal gas (as $n_{\rm crit}\propto{T}^{1/2}$).
However the high values of critical density are consistent with the coronal
lines arising in a relatively high density gas, such as that confined to the 
``equatorial'' region, i.e. the region confined to the orbital
plane (see below).

The fall and rise of some of the coronal line fluxes as the 
excitation mechanism switches from photionisation+shock, to
shock only, is seen around day~15 in Fig.~\ref{coronal_lc}.
A similar dip in coronal line fluxes, around day~70, was reported 
for the 2006 eruption of RS~Oph by \cite{evans07b}, who attributed 
this behaviour to the breakout of the ejecta from the RG wind.
However the relative timescales of the SSS phase in \rnsgr\ and 
RS~Oph ($\sim20$~days and $\sim60$~days respectively) are similar 
to the times at which the ``dips'' occur in their respective coronal 
line light curves, suggesting that the duration of the SSS phase 
is a more likely explanation than breakout. This is 
consistent with a comparison of the distance travelled by the outer 
edge of the RG wind since the 1990 eruption to the distance travelled 
by the decelerating ejecta in the 2019 eruption. 
\cite{kaminsky22} suggest a wind velocity of 17\vunit\ for the 
RG in \rnsgr; this is similar to that seen in field RGs 
\citep[see, e.g.,][]{wood16}. The outer edge of the RG wind 
was therefore at $\sim1.6\times10^{15}$~cm at the 
time of our observations, while in time $t$ the 2019 ejecta will 
have travelled $\sim8.6\times10^{13}\,(t/5~\mbox{days})^{0.25}$~cm,
assuming that the ejecta velocity varies as $t^{-0.75}$ 
(see Fig.~\ref{H_decel}). At the time of our last outburst 
observation ($t=46.35$~days), this is $1.5\times10^{14}$~cm. 
Clearly the ejecta were a long way from reaching the edge of 
the wind. The SSS hypothesis for 
the dips can obviously be tested when other RNe with RG 
secondaries erupt, in particular the 2021 eruption of RS~Oph 
\citep{amorim21} and the imminent eruption of T~CrB \citep{luna20}.

\section{Discussion}
\label{disc}
The usual paradigm for a nova eruption that occurs in a system 
containing a RG is that the ejected material collides with and
collisionally shocks the RG wind.
This is commonly evidenced by the appearance of coronal line emission
in the optical and IR, X-ray emission and non-thermal radio emission,
and the deceleration of the ejecta 
\citep[see contributions in][which reviews the 2006 eruption of RS~Oph]{evans-aspc}.

In the 2019 RN eruption of \rnsgr, all of these ingredients are present, but
here we consider the implications of the NIR data. 
We summarise our findings as follows.
\begin{enumerate}
\item On two dates (2019 September 8 and 20), 
there is evidence for a cool ($\sim7000$~K) gas that
produced a free-free and free-bound continuum;
 \item the H recombination and \pion{He}{i} lines consist of narrow 
 components superposed on broad pedestals;
 \item the narrow components become progressively narrower in the H 
 recombination lines, but the broad components remain
 stubbornly broad ($\mbox{FWZI}\simeq8000$\vunit) throughout our observations.
 \cite{munari19b} also noted that the width of the broad pedestals
 in optical emission lines had not changed during the first $\sim15$~days.
 Both their data and ours imply 
 that little deceleration of the material responsible for the 
 broad pedestal emission occurred during this time;
 \item the narrow components of the H and He lines persist for the duration
 of our observations, although they weaken slightly with time;
 \item the fluxes in the broad components decline substantially over the
 period of our observations;
 \item coronal line emission is present from day~5.11; the coronal lines 
 are narrow throughout the period of our observations;
 \item  the temperature of the coronal gas, as determined 
 from silicon and sulphur lines, is $\sim9.3\times10^5$~K;
 \item abundance ratios relative to Si of five elements, 
 as determined from the coronal lines, are broadly similar to solar.
\end{enumerate}

The progression of a nova eruption in a system containing a WD with 
an accretion disc, and a RG with a wind, has been considered in 
detail by \cite*{booth16} in the context of the RN RS~Oph,
which superficially resembles \rnsgr\ in many ways.
They find that, in the inter-eruption period, the interaction of the 
RG wind with the binary causes the majority of the wind to be confined 
to a circumbinary disc close to the orbital plane of the binary, 
and that the disc may be somewhat inclined to the binary plane 
\citep[as discussed by][]{theuns93}.
Observational evidence for enhanced density in the 
equatorial region for the case of RS~Oph has been given by \cite{ribeiro09}.
Presumably the ``old'' silicate dust seen in the environment of
RS~Oph \citep{evans07c,woodward08,rushton22}, and the cool
dust in \rnsgr\ \citep{kaminsky22}, reside in such a structure.

When the RN eruption occurs \citep[see][for details]{booth16}, 
the presence of the accretion disc restricts the flow in the orbital 
plane and imposes bipolar geometry on the nova ejecta: the ejected
material escapes perpendicular to the orbital plane
at $\sim4000$\vunit. This is the same order of ejection velocity that 
we see in the broad emission line components in \rnsgr\ (the polar component).
Our observed FWZI of $\sim8000$\vunit\ translates to an
ejection velocity of $V=10700$\vunit\ for an inclination of $68^\circ$
\citep{mikolajewska21}. Material ejected in the binary plane
(the ``equatorial'' component) encounters the accumulated debris from
the RG wind, previous eruptions and the common envelope phase.

As noted in subsection~\ref{coronal_s}, there is evidence
that the equatorial ejecta were encountering significant 
resistance from near the time of outburst.
This behaviour is similar to that seen in the 2014 eruption of the 
RN V745~Sco, when the ejecta began sweeping up circumstellar material from
the outset \citep{banerjee14}. 
The case of RS~Oph is more complex, in that decelaration
did not start until $\sim$~day~5 in the 2006 eruption \citep*{das06},
but much earlier ($\sim3.9$~days) in its 2021 eruption \citep{pandey22}.

While we must be wary of extrapolating and of taking the fit too 
literally, it is intriguing that the $V\propto{t}^{-0.74}$
dependence for the velocities corresponding to the HWHM of
the narrow components (see Section~\ref{hwhm}) extrapolates to 
$V\sim10000$\vunit\ at $t=0.1$~d, close to the velocity of 
the polar component. We might infer from this that all material was 
ejected at $\sim10000$\vunit\ in the TNR, ejecta in the polar 
direction moving without impediment, ejecta in the equatorial
direction being severely decelerated from the outset, to
$V\sim100$\vunit.

The binary separation in \rnsgr\ is 2.2~AU \citep{mikolajewska21}.
Material ejected in the 2019 eruption, decelerating 
as $V\propto{t}^{-0.75}$, would reach the RG in $\sim2$~hours.
To a good approximation we may therefore suppose that the RN
ejecta and the RG wind are co-centric, and that the ejecta
run spherically symmetrically into the RG wind. We take the wind
mass-loss from the RG to be $10^{-7}$\Msun\,yr$^{-1}$ 
\citep[see, e.g.,][]{espey08,banerjee14}, at velocity 17\vunit\
\citep{kaminsky22}. The wind mass swept up by the ejecta in time $t$, 
decelerating as discussed above, is $\simeq1.2\times10^{-7}\:{t}^{0.25}$\Msun, or 
$\simeq1.8\times10^{-7}$\Msun\ in 5~days.

Under the assumptions that led to Equation~(\ref{blob}), the 
ejected mass $M_{\rm ej}$ in the polar direction 
$M_{\rm ej}\propto\varpi{V}\delta{t}$, where $\varpi$ is 
the solid angle of the cone into which the material is ejected.
The ejected mass depends on $\varpi$ and $D$ as 
$M_{\rm ej}\propto\varpi^{1/2}D$; we assume that $D=9$~kpc \citep{mikolajewska21}. 
Using the dereddened fluxes of the broad components of the 
Pa\,$\alpha$, Pa\,$\beta$ and Br\,$\gamma$ lines,
we can estimate the ejected mass if we assume Case~B, with an
electron temperature of $T_e=2\times10^4$~K. We take an arbitrary 
value of $\varpi$ corresponding to an apex angle for the conal ejection
of $5^\circ$, a highly collimated ejection.

\begin{figure}
 \centering
 \includegraphics[width=7.5cm]{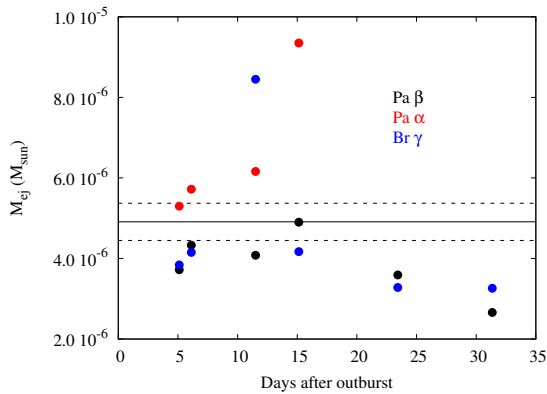}
 \caption{Total ejected mass in the polar components determined
 as described in the text. The horizontal line is the mean
 value, the broken lines are $\pm1$~standard error of the mean.\label{blobf}}
\end{figure}

The resuls are shown in Fig.~\ref{blobf}, which includes the H~mass
in both polar components. Considering the simple 
nature of the ``model'', and the uncertainties in disentangling the 
broad and narrow line components, the individual masses are, with 
two exceptions (Br\,$\gamma$ on day~11.51 and Pa\,$\alpha$ on day~15.13), 
satisfyingly consistent. The mean value is $4.8[\pm0.5]\times10^{-6}$\Msun\ 
and, as this is the mass in the form of H only, it represents a lower limit. 
The corresponding kinetic energy is $5.4\times10^{38}$~J. 
While this is an estimate of the mass in the polar
ejecta, we might suppose that the mass in the equatorial ejecta is not
dissimilar, and if so, this mass is greater than our estimate of the RG
wind mass swept up by the ejecta.
Our estimated mass is similar to that ejected 
in other long-orbital period RNe \citep[$3-4\times10^{-6}$\Msun;][]{anupama08}.

With no complications arising from the RG wind,
the line emission from the polar component should 
be determined by photoionisation only. For a TNR on the surface of a WD 
with mass close to the Chandrasekhar limit, the mass of helium in 
the ejecta is expected to be $\gtsimeq0.6$ times the mass of hydrogen
depending on the degree of mixing of WD core
and accreted material; see, e.g.,][]{starrfield20}, 
in contrast to the solar value of $\sim0.3$. 
To test this, we estimate
the He/H ratio using the prescription of \cite{krautter84} which,
for reasons discussed by those authors, gives a lower limit on
$N(\mbox{He}^+)/N(\mbox{H}^+)$. We use the measured fluxes in
the broad components of the \pion{He}{i} 2.058\mic\ and Br\,$\gamma$
lines; the wavelengths are close enough that uncertainties in
reddening and the need to deconvolve the lines from the 
instrumental resolution are not important. For days 5.11--31.34
(the broad components are too weak on day 46.35), we find a mean 
value for the lower limit on $N(\mbox{He}^+)/N(\mbox{H}^+)$ of 
$0.9\pm0.2$, implying that the He/H ratio by number is at least unity.
This is an order of magnitude higher than the solar
H/He ratio \citep[0.082 by number;][]{asplund21} and is consistent
with the expectation for a TNR on a massive WD.

However the large wind mass close to the orbital plane,
the accumulation of material lost by the RG wind over several
orbits and possibly some material remaining from the common envelope 
phase, cause the nova ejecta to decelerate in this direction, 
to $\sim100-200$\vunit; the ejecta sweep up the accumulated wind. 
In the $\sim29$~years since the 
1990 eruption, the accumulated wind in the circumstellar 
environment would have a mass $\sim2.9\times10^{-6}$\Msun\ 
if the RG has mass-loss rate $\sim10^{-7}$~\Msun\,yr$^{-1}$. 
This is comparable with the mass we estimated for the polar component
(see above). Whereas the high velocity (polar)
material is representative of material ejected in the TNR, 
the decelerated material, which encounters the RG wind,
is expected to have a composition that is close to that of the RG 
\citep{booth16}. The coronal lines must arise in the equatorial 
component not yet reached
by the RN ejecta, as they are persistently narrow and,
unlike the H recombination lines, show no evidence of deceleration.
The abundances we find in the coronal region (see Table~\ref{abund}), 
which must reside in the equatorial region, are approximatey solar;
this is consistent with the scenario outlined by 
\cite[][but note our earlier caveat regarding Galactic abundance gradients]{booth16}.

The cool material, evidenced by the presence of the Paschen and 
Brackett discontinuities (see Fig.~\ref{ff-bf}), presumably also 
lies in the equatorial region, but in a region where the electron
density is $<4\times10^7$~cm$^{-3}$ \citep[][see Section~\ref{2019}]{munari19b}.
Such a region is likely located on the outer edge of the equatorial
region, where the electron density is likely to be lowest.
This material was ionised during the eruption, but the 2019 ejecta have
not yet reached it, and it has not recombined.

\begin{figure}
\centering
 \includegraphics[width=8cm]{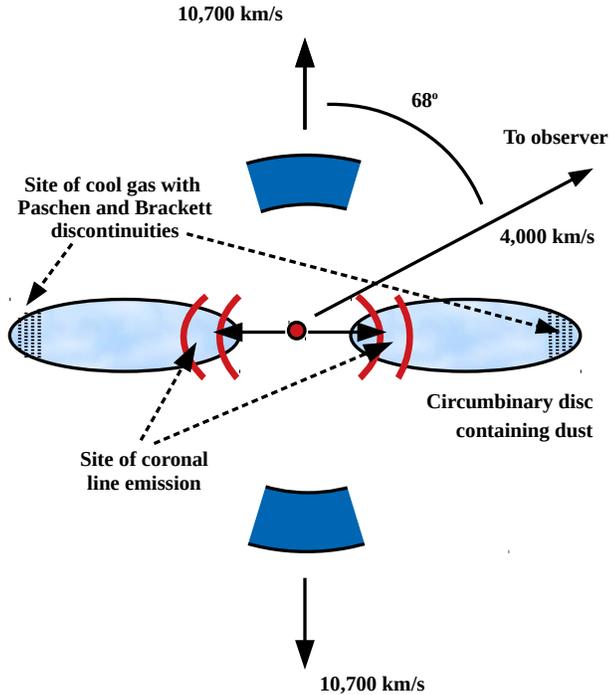}
 \caption{Sketch of the binary environment as suggested by the IR spectra.
 The binary is located at the small red dot at the centre, the circumbinary
 disc is represented by the pale blue ellipses. The cool gas responsible for 
 the Paschen and Brackett discontinuities is located in the outer,
 low density, portions of the circumbinary disc. The dark blue regions
 depict the polar ejecta, travelling at 10700\vunit.
 The red arcs represent the decelerating equatorial ejecta. 
 The direction to the observer is indicated.
 \label{toon}}
\end{figure}

A cartoon of the likely circumbinary environment of \rnsgr\ 
during euption is shown in Fig.~\ref{toon}.

\section{Conclusion}

We have presented NIR spectroscopy of the 2019 eruption
of the recurrent Nova \rnsgr. We conclude that
\begin{enumerate}
 \item the evolution of the emission lines suggest two distinct
regions of ejected gas, one that results from the escape of ejecta 
 at high velocity perpendicular
 to the binary plane (the polar component), and another that results 
 from the interaction of both radiation from the nova and its high velocity
 ejecta with the accumulated wind of the RG and the detritus from previous
 RN eruptions (the equatorial component), resulting in rapid deceleration 
 of the ejecta;
 \item the coronal line emission arises in the material that is
 confined to the equatorial component in the plane of the binary;
 \item the decline in the fluxes of the broad components of the 
 H and \pion{He}{i} emission lines suggests that the material was 
 ejected in a brief pulse of about 3.3~days duration;
 \item the lower limit on the He/H ratio in the high-velocity material
 is consistent with the origin of this material in a TNR on 
 the surface of a massive WD;
 \item the behaviour of the coronal line fluxes suggest that there were 
 two coronal phases, an earlier phase when the coronal gas was both 
 shock-ionised and photoionised, 
 and a later phase when the coronal gas was truly coronal, in that the gas
 was collisionally heated and the transition upper levels were populated by
 electron impact;
 \item the temperature of the coronal gas was $9.3\times10^5$~K;
 \item the metal abundances are consistent with solar and RG
values, and inconsistent with those expected from a TNR on a massive WD.
 \item our data broadly support the scenario detailed by \cite{booth16}.
\end{enumerate}

\section*{Acknowledgements}

We thank the referee for their comments
which have helped to improve the paper.

The Gemini observations were made possible by awards of
Director's Discretionary Time for programmes
GS-2019B-DD-102 and GN-2019B-DD-104.
The international Gemini Observatory is a program of 
NSF's NOIRLab, which is managed by the Association of Universities 
for Research in Astronomy (AURA) under a cooperative agreement 
with the National Science Foundation, on behalf of the Gemini 
Observatory partnership: the National Science Foundation 
(United States), National Research Council (Canada), 
Agencia Nacional de Investigaci\'{o}n y Desarrollo (Chile), 
Ministerio de Ciencia, Tecnolog\'{i}a e Innovaci\'{o}n (Argentina), 
Minist\'{e}rio da Ci\^{e}ncia, Tecnologia, 
Inova\c{c}\~{o}es e Comunica\c{c}\~{o}es (Brazil), 
and Korea Astronomy and Space Science Institute (Republic of Korea).

Additional data presented in 
this paper were obtained partly under IRTF programme 2020A-010.
The Infrared Telescope Facility is operated by the
University of Hawaii under contract 80HGTR19D0030 with the 
National Aeronautics and Space Administration.

CEW acknowledges partial support from NASA grant 80NSSC19K0868.
DPKB is supported by a CSIR Emeritus Scientist grant-in-aid and 
is being hosted by the Physical Research Laboratory, Ahmedabad. 
RDG was supported by the United States Airforce.

We gratefully acknowledge the contributions of
Terry Bolhsen and Umberto Sollecchia to the ARAS database.

We acknowledge with thanks the variable star observations from the 
AAVSO International Database contributed by observers worldwide and 
used in this research.

\section*{Data availability}
The raw data in this paper are available from the Gemini Observatory
Archive, https://archive.gemini.edu/, and from the IRTF archive,
http://irtfweb.ifa.hawaii.edu/research/irtf\_data\_archive.php






\bsp	
\label{lastpage}
\end{document}